\documentclass[aps,prb,twocolumn,superscriptaddress]{revtex4-1}
\pdfoutput=1
\usepackage[latin9]{inputenc}
\usepackage{graphicx,epsf}
\usepackage{amssymb}
\usepackage{amsmath}
\usepackage{color}
\usepackage{bm}
\usepackage[caption=false]{subfig}
\usepackage{placeins}

\newcommand{\w}{\omega}

\newcommand{\loc}{{\rm loc}}
\newcommand{\imp}{{\rm imp}}
\newcommand{\hyb}{{\rm hyb}}
\newcommand{\const}{{\rm const}}

\RequirePackage[
   hyperindex,colorlinks,bookmarksnumbered,
   plainpages=true, pdfstartview=FitH,bookmarks=false
]{hyperref}
\hypersetup{linkcolor=blue,urlcolor=blue,citecolor=blue}

\definecolor{darkgreen}{rgb}{0,0.5,0}
\definecolor{darkblue}{rgb}{0,0,0.5}
\definecolor{purple}{rgb}{0.35,0,0.35}
\definecolor{orange}{rgb}{1,0.5,0}
\definecolor{todocolor}{rgb}{1,0,0}
\definecolor{jvdcolor}{rgb}{0,0,1}

\graphicspath{{./}{./figures/}}

\begin{document}
\title{
Mott quantum criticality in the one-band Hubbard model: \\
Dynamical mean-field theory, power-law spectra, and scaling
}

\author{Heike Eisenlohr}
\affiliation{Institut f\"ur Theoretische Physik and W\"urzburg-Dresden Cluster of Excellence ct.qmat, Technische Universit\"at Dresden,
01062 Dresden, Germany}
\author{Seung-Sup B. Lee}
\affiliation{Faculty of Physics, Arnold Sommerfeld Center for Theoretical Physics, Center for NanoScience, and Munich Center for Quantum Science and Technology, Ludwig-Maximilians-Universit\"{a}t M\"{u}nchen, Theresienstra{\ss}e 37, 80333 M\"{u}nchen, Germany}
\author{Matthias Vojta}
\affiliation{Institut f\"ur Theoretische Physik and W\"urzburg-Dresden Cluster of Excellence ct.qmat, Technische Universit\"at Dresden,
01062 Dresden, Germany}

\date{\today}

\begin{abstract}

Recent studies of electrical transport, both theoretical and experimental, near the bandwidth-tuned Mott metal-insulator transition have uncovered apparent quantum critical scaling of the electrical resistivity at elevated temperatures, despite the fact that the actual low-temperature phase transition is of first order. This raises the question whether there is a hidden Mott quantum critical point.
Here we argue that the dynamical mean-field theory of the Hubbard model admits, in the low-temperature limit, asymptotically scale-invariant (i.e. power-law) solutions, corresponding to the metastable insulator at the boundary of metal-insulator coexistence region, which can be linked to the physics of the pseudogap Anderson model. While our state-of-the-art numerical renormalization group calculations reveal that this asymptotic regime is restricted to very small energies and temperatures and hence difficult to access numerically,
we uncover the existence of a wide crossover regime where the single-particle spectrum displays a \emph{different} power law. We show that it is this power-law regime, corresponding to approximate local quantum criticality, which is continuously connected to and responsible for the apparent quantum critical scaling above the classical critical end point.
We connect our findings to experiments on tunable Mott materials.
\end{abstract}

\maketitle


\section{Introduction}

The interaction-driven metal-insulator transition is a paradigmatic example of strong correlation effects, and yet remains poorly understood. Significant theoretical progress has been made using the framework of dynamical mean-field theory (DMFT) which approximates the electronic self-energy as local in space -- an approximation which becomes exact in the limit of infinite space dimensions. Within DMFT, the bandwidth-tuned Mott transition of the half-filled one-band Hubbard model has been established\cite{Georges93,Rozenberg94,Georges96,Bulla99,Bulla01} to be discontinuous at low temperature $T$, with a first-order transition line terminating at a classical critical endpoint located at a temperature $T_c$ which is a small fraction of the bandwidth. This phenomenology agrees with a variety of experimental results, obtained from compounds with metal-insulator transitions driven by (hydrostatic or chemical) pressure.

Given that the low-temperature Mott transition is discontinuous,\cite{zerot_note} it came as a surprise when apparent quantum critical scaling of the electrical resistivity was reported at elevated temperatures above the critical endpoint. Such scaling was first found in numerical DMFT studies \cite{Dobro11, Dobro13} of the half-filled Hubbard model. This triggered corresponding experiments in quasi-2d organic salts \cite{Kanoda15} where then resistivity scaling in agreement with the theoretical predictions was found.

Conceptually, the observed quantum critical behavior is puzzling, as it suggests the existence of an underlying Mott quantum critical point (QCP) which, however, does not appear in the DMFT phase diagram. In principle, the QCP might be uncovered upon varying another tuning parameter $x$ such as doping, so that $T_c(x)$ vanishes at a particular $x$ leading to a quantum critical endpoint. However, it was found numerically \cite{Dobro15} that 
tuning of the chemical potential does not change the first-order nature of the transition, i.e. $T_c$ remains finite, while quantum critical scaling persists.
In addition, the fact that two-dimensional experimental systems display behavior similar to DMFT indicates that momentum dependencies are weak and the behavior is driven by local correlations, speaking against full space-time quantum criticality.

\begin{figure}
 \includegraphics[width=\columnwidth]{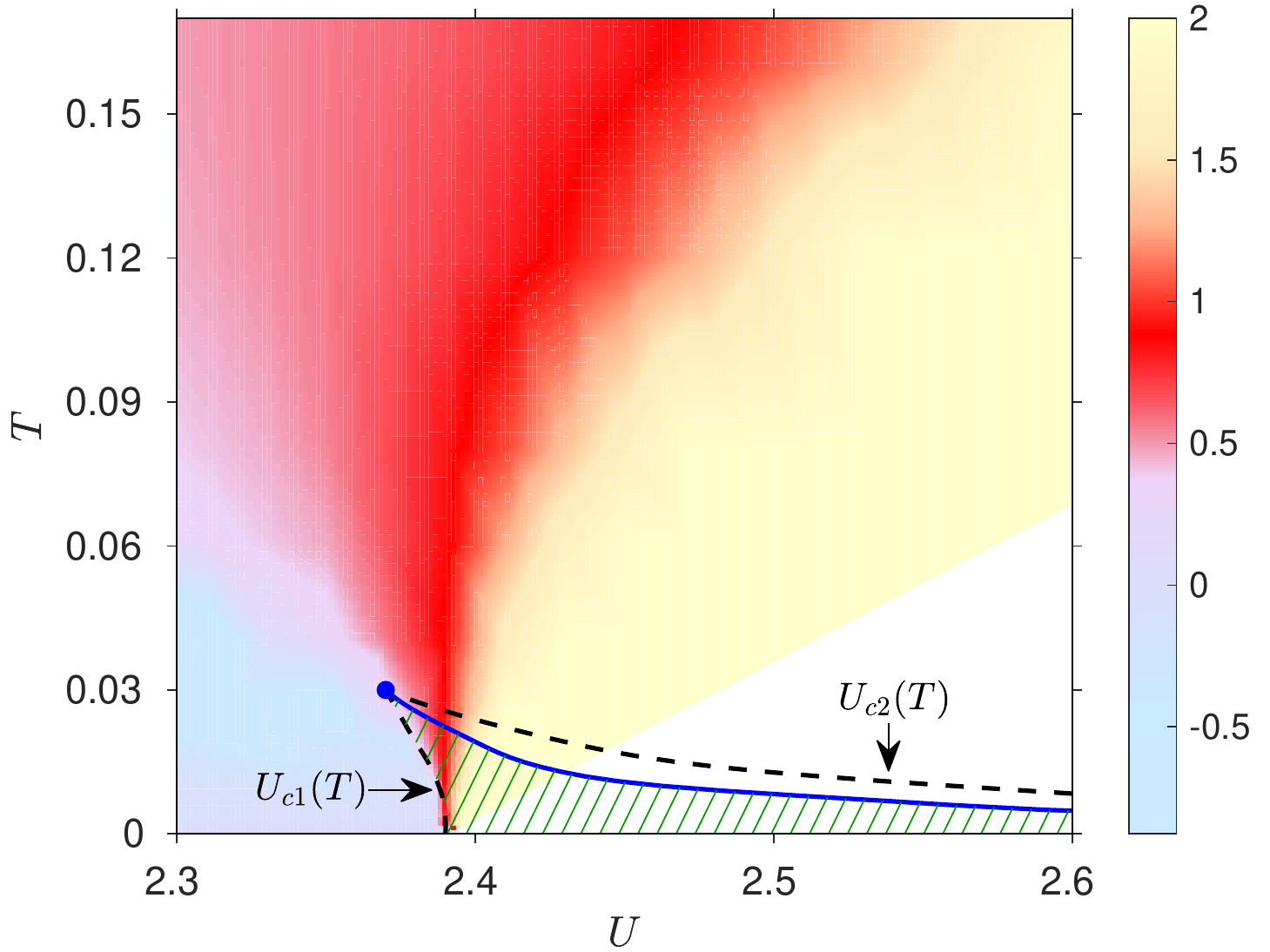}
 \caption{
 DMFT phase diagram of the half-filled Hubbard model as function of interaction strength $U$ and temperature $T$, both measured in units of the half-bandwidth $W$, with the first-order transition line (blue), the critical endpoint (dot), and the boundary of the coexistence region (black dashed).
  The colormap shows the exponent of the spectral function, $r_T = d\ln A(\w) / d\ln\w$, at $\w=2 \,T$. A wide regime with $r_T=0.7 \ldots 0.8$ (red) is visible which extends down to very low $T$, corresponding to the crossover power law which controls the approximate quantum critical behavior.
  In the coexistence region below $T_c$, data for the insulating solution is shown, which is metastable in the green hatched region. Data have been cut off deep in the insulator where $A(\omega=2T)<10^{-8}$ (white). For details see Sec.~\ref{sec:hybT}.}
 \label{fig:fancyphasediagram}
\end{figure}

The purpose of this work is to shed light onto this puzzle. To this end, we study the one-band Hubbard model within DMFT both analytically and numerically. First, we show that DMFT admits scale-invariant solutions in the limit $T\to 0$ where the single-particle spectrum displays power-law frequency dependence and which do {\em not} correspond to a QCP of the underlying impurity problem, but are solely driven by DMFT self-consistency. We conjecture that a metastable\cite{unstable_note} power-law solution is indeed realized at the small-$U$ boundary of the metal-insulator coexistence region.
Second, using extensive numerical simulations with Wilson's numerical renormalization group (NRG) technique as impurity solver, we find that the corresponding asymptotic regime is restricted to very low energies and temperatures. Over a wide range of intermediate energies and temperatures we instead find a \emph{crossover} spectral power law which exists both in the insulating DMFT solution below $T_c$ and in the metal-insulator crossover region above $T_c$, Fig.~\ref{fig:fancyphasediagram}. Studying the electrical resistivity, we are able to extend the apparent quantum critical scaling found above $T_c$ down to low $T$ and to connect it to the spectral power law. We hence establish an instance of approximate local quantum criticality.
We also comment on the role of Widom lines in the phase diagram and discuss implications beyond DMFT as well as pertinent experiments.


\subsection{Summary of previous results}

The DMFT solution of the one-band Hubbard model has been studied in numerous papers in the past. At low $T$, a first-order metal-insulator transition is found as function of $U/t$, with metallic and insulating solutions coexisting between $U_{c1}$ and $U_{c2}$.\cite{zerot_note,Georges93,Rozenberg94,Georges96,Bulla99,Bulla01}

Refs.~\onlinecite{Dobro11, Dobro13} calculated the electrical resistivity $\rho$ of the one-band Hubbard model within DMFT, using both iterated perturbation theory (IPT) and continuous-time Quantum Monte Carlo (CT-QMC) impurity solvers. At fixed electronic bandwidth, the resistivity $\rho(U,T)$ as function of Hubbard interaction $U$ and temperature $T$ was found to display scaling behavior, i.e.,
\begin{equation}
\label{eq:dobroscale}
\frac{\rho(\delta U,T)}{\rho_c(T)} = f_\pm\left(\frac{T}{|\delta U|^{\nu z}}\right)
\end{equation}
where $\delta U = U - U^\ast(T)$ is the interaction measured relative to the quantum Widom line $U^\ast(T)$, \mbox{$\rho_c(T)=\rho(\delta U\!=\!0, T)$} is the resistivity on the Widom line, and $f_\pm$ are scaling functions for $\delta U \gtrless 0$.
The importance of the quantum Widom line as a reference line for the scaling analysis was stressed in Refs.~\onlinecite{Dobro11, Dobro13}; it represents the continuation of the first-order transition line at low $T$ into the crossover regime $T>T_c$. We note that standard quantum critical scaling would define $\delta U = U-U_{\rm cr}$ with $U_{\rm cr}$ being the zero-temperature location of the QCP, corresponding to $U^\ast(T)=\rm const$.
In Refs.~\onlinecite{Dobro11, Dobro13} the resistivity data were found to display a good scaling collapse according to Eq.~\eqref{eq:dobroscale} in a temperature regime $2 \, T_c < T < 4 \, T_c$, with the exponent $\nu z \approx 0.6$ from IPT and slightly larger for CT-QMC.

The analysis of the experimental resistivity data\cite{Kanoda15} on organic triangular-lattice Mott compounds also employed Eq.~\eqref{eq:dobroscale}, with pressure $P$ corresponding to $t/U$ and $P^\ast(T)$ chosen as the inflection point of $\ln \rho(P,T)$ at fixed $T$. For the materials \mbox{$\kappa-($ET$)_2 $Cu$ [$N$($CN$)_2] $Cl}, \mbox{$\kappa-($ET$)_2 $Cu$_2 ($CN$)_3$} and \mbox{EtMe$_3$Sb$[$Pd$($dmit$)_2]_2$} a good scaling collapse was found again in the temperature range $2 \, T_c < T < 4 \, T_c$, with values for $\nu z$ ranging from $0.5$ to $0.7$.
Universality among various Mott compounds was pointed out in Ref.~\onlinecite{Dobro17}, where optical spectroscopy was used to determine microscopic energy scales which allowed a mapping of rescaled phase diagrams.

Finally, the local spin susceptibility was reported from DMFT to show $\omega/T$ scaling at fixed $U$ in the regime above the critical endpoint. \cite{McKenzie17} This analysis used a temperature-independent reference line $U^\ast(T)$.

We note that critical Mott-like behavior has been analyzed for the Falicov-Kimball model in Refs.~\onlinecite{Haldar16,Haldar17}; however, this model displays a continuous metal-insulator transition as $T\to 0$ and explicitly breaks spin rotation symmetry.


\subsection{Outline}

The remainder of the paper is organized as follows:
Sec.~\ref{sec:model} gives a quick introduction into the Hubbard model, its treatment using DMFT, and the NRG technique. Sec.~\ref{sec:ana} discusses the possibility of DMFT solutions with power-law frequency dependence of spectra and self-energies; such scale-invariant solutions are a prerequisite for quantum criticality in DMFT.
In Secs.~\ref{sec:numzerot} and \ref{sec:numfinitet} we show our numerical DMFT results, first in the limit of zero temperature and then for finite $T$. Sec.~\ref{sec:numfinitet} includes a discussion of resistivity scaling and of Widom lines.
A discussion of broader implications of our results closes the paper.
Numerical results for other bare densities of states (i.e. other lattices) are shown in the appendix.


\section{Hubbard model and DMFT}
\label{sec:model}

We consider the one-band Hubbard model at half-filling. The Hamiltonian is given by
\begin{align}
 H= -t \sum_{\langle i j \rangle\sigma} c_{i\sigma}^\dag c_{j\sigma}^{\phantom \dag} + U \sum_i \left(n_{i \uparrow} - \frac 1 2 \right) \left( n_{i \downarrow} - \frac 1 2 \right)
\end{align}
where $i,j$ label lattice sites in real space. We define the local Green's function as
$ G_{\loc}(\tau) = - i \langle \mathcal{T} c_{i \sigma}^{\phantom \dag}(\tau) c_{i \sigma}^{\dag}(0) \rangle $.
For most of the following, we focus on the case with semicircular density of states (DOS) $\rho(\epsilon) = 2 \sqrt{W^2-\epsilon^2}/\pi W^2$, corresponding to the Bethe lattice in infinite dimensions, and use the semi-bandwidth $W=2t$ as our unit of energy. Other choices of bare DOS are discussed in the appendix.


\subsection{Dynamical mean-field theory}

In the framework of DMFT the Hubbard model is mapped onto an effective impurity model by neglecting non-local contributions to the self-energy -- an approximation which becomes exact in the limit of infinite lattice coordination.\cite{Georges96} The effective action of this single-impurity Anderson model is
\begin{align}
 S_{\rm eff}=&-\int_0^\beta \!d\tau d\tau^\prime \sum_\sigma c_{0\sigma}^\dag(\tau) \tilde G_0^{-1}(\tau - \tau^\prime)  c_{0\sigma}^{\phantom \dag}(\tau^\prime)
 \nonumber \\
  &+ U \int_0^\beta \!d\tau \; (n_{0 \uparrow} (\tau) - \frac 1 2) (n_{0 \downarrow}(\tau) - \frac 1 2)
\end{align}
where $\tilde G_0(i \omega_n)= i \omega_n +\mu -\Gamma(i \omega_n)$ is the non-interacting impurity Green's function, and $\mu=0$ for the present particle-hole-symmetric case. The hybridization function $\Gamma(i \omega_n)$ determines the coupling of the impurity to the bath and plays the role of a dynamical mean field. The crucial step is to fix this hybridization self-consistently by demanding equality of the impurity Green's function with the local lattice Green's function:\cite{Georges96}
\begin{align}
 \tilde G_0(i \omega_n) = i \omega_n + \mu + G_{\loc}^{-1}(i \omega_n) - R[G_{\loc}(i \omega_n)]\,.
\end{align}
Here $R$ is the inverse Hilbert transform of the bath DOS, defined by
\begin{equation}
 D(z) = \int_{-\infty}^\infty d \epsilon \frac{\rho(\epsilon)}{z-\epsilon},~~~
 R[D(z)] = z.
\end{equation}
In the case of the semicircular DOS, the inverse Hilbert transformation can be evaluated analytically to give
\begin{align}
 R[D(z)]= \frac{W^2}{4} D(z) + \frac{1}{D(z)},
\end{align}
such that the self-consistency equation simplifies to
\begin{align}
 \Gamma(i \omega_n) = \frac{W^2}{4} G_{\loc}(i \omega_n),
 \label{eq:BetheSC}
\end{align}
i.e. the hybridization function is proportional to the local lattice Green's function.
The self-consistency condition can also be formulated in terms of the interaction-induced self-energy $\Sigma(i \omega_n)$, defined by
\begin{align}
  \Gamma(i \omega_n) = \frac{W^2}{4} \int d \epsilon \frac{\rho(\epsilon)}{i \omega_n + \mu - \Sigma(i \omega_n)-\epsilon} \,.
\end{align}
In the following we will primarily consider the bath spectral function on the real frequency axis \mbox{$A(\omega) = - \Im \Gamma(\omega+i0^+)/\pi$}.

Within DMFT the temperature-dependent dc resistivity is calculated from the zero-frequency optical conductivity as \mbox{$\rho = 1/\sigma(\omega\!=\!0)$} and is given in units of the largest possible metallic conductivity \mbox{$\rho_{\rm Mott} = \hbar a/e^2$}.
The frequency-dependent optical conductivity is given by
\begin{align}
  \sigma(\omega)= & 2 \pi \frac{e^2}{\hbar a} \int_{-\infty}^{\infty} \! d\omega^\prime \; \frac{n_F(\omega^\prime)-n_F(\omega^\prime+\omega)}{\omega} \nonumber \\
  &\times \int_{-\infty}^{\infty} \! d\epsilon \; \rho(\epsilon) \left( v(\epsilon) \right)^2  A(\epsilon, \omega^\prime) A(\epsilon, \omega^\prime+\omega)
\label{eq:dmftsigma}
\end{align}
with $n_F(\w)$ as the Fermi distribution
and $v(\epsilon) = \sqrt{W^2 - \epsilon^2}/\sqrt{3}$ the velocity, here for the Bethe lattice in \mbox{$z \rightarrow \infty$}.\cite{Georges96, Merino, georges13}
The spectral function $A(\epsilon, \w)= - \Im\left( G(\epsilon, \w) \right)/\pi$ is calculated from the DMFT Green's function
$G(\epsilon, \w) = \left[(G_0(\epsilon, \w)^{-1} - \Sigma(\w) \right]^{-1}$. The non-interacting Green's function $G_0(\epsilon, \w) = 1/(\w - \epsilon + \mu)$ depends on the one-particle energy $\epsilon$, which is used here as a quantum number replacing momentum. Importantly, current vertex corrections are absent in DMFT \cite{Georges96} such that single-particle spectrum and conductivity are directly related.


\subsection{Numerical renormalization group}

For solving the impurity problem within DMFT we use the NRG technique as pioneered by Wilson \cite{wilson_rev_1975} which is particularly suitable to access the physics at low energies and temperatures; a thorough introduction to the method is given in Ref.~\onlinecite{bulla_rev_2008}. Its accuracy and its formulation on the real frequency axis favor NRG over the Quantum Monte Carlo and iterated perturbation theory impurity solvers used in previous studies \cite{Dobro11,Dobro13,Dobro15} of this regime. We utilize the tensor-network formulation of NRG, \cite{Weichselbaum12a} and make use of the so-called self-energy trick,\cite{Bulla98} as well as z-averaging\cite{Oliveira94,*Oliveira97} with appropriate logarithmic discretization \cite{ZitkoPruschke09, SeungSup19} for $\Lambda=2$ and $n_z=4$. The full density matrix as constructed from discarded states is used to calculate spectral functions (fdm-NRG),\cite{Weichselbaum07} and the non-abelian SU(2) symmetries of charge and spin conservation are taken into account to reduce the size of matrix blocks.\cite{Weichselbaum12b}  In each NRG iteration we truncate the Hilbert space to maximally $3000$ multiplets or, at later iterations, by applying an energy cutoff $E_\mathrm{trunc}=15$.

The broadening of discrete spectral data has to be done carefully in order to obtain gapped insulating solutions at low $T$. We employ the adaptive broadening scheme proposed in Ref.~\onlinecite{Weichselbaum16}. It uses the sensitivity of a spectral feature to an infinitesimal z-shift to determine its optimal broadening width for logarithmic broadening kernels and then applies a linear broadening kernel with width $\sigma \sim T$ to all frequencies to avoid artifacts which would arise from a switching of broadening protocols on the scale $\omega \sim T$. This scheme was shown to give very accurate results for the Hubbard model.\cite{Lee17} In the nomenclature of Ref.~\onlinecite{Weichselbaum16}, for logarithmic broadening we use $\alpha=1.5$ and limit the broadening width to $\ln \Lambda/5 \geq \sigma_{ij} \geq \ln \Lambda /15$.  For the linear broadening we use $\gamma=\min(T/10, 0.001)$.

As a convergence criterion for the DMFT self-consistency loop we use $\max_\omega |A_{\text{in}}(\omega)-A_{\text{out}}(\omega)| < 10 ^{-4}$. Close to $U_{c1}$ this is not sufficient to distinguish a metastable insulating from a non-converged metallic solution, so the convergence threshold was lowered. To achieve convergence between $10$ and $200$ DMFT iterations are used.


\section{Scale-invariant DMFT solutions and the pseudogap Anderson model}
\label{sec:ana}

Quantum criticality requires the existence of a scale-invariant state at zero temperature, i.e., a gapless state with power-law correlations in both space and time.\cite{Sachdevbook} Due to the local approximation to the self-energy on which DMFT is based, only correlations in time are relevant.
We therefore start with analytical considerations on the possibility of scale-invariant solutions of the Hubbard-model DMFT equations at zero temperature, as the existence of such solutions appears as a necessary condition for DMFT-based Mott quantum criticality.

Within DMFT, a scale-invariant solution must be characterized by a local single-particle spectral function with power-law behavior, $G_\loc(\w) \propto |\w|^r$, at low energies. As we are looking for a solution corresponding to the transition between a metal ($r=0$) and an insulator (formally $r\to\infty$), we can expect $0<r<\infty$.
For non-integer $r$ this implies a self-energy which is dominated by $\Sigma(\w) \propto |\w|^{-r}$ (more precisely,\cite{KramersKronig_note} $\Im \Sigma(\w) \propto |\w|^{-r}$ and $\Re \Sigma(\w) \propto \text{sgn}(\w) |\w|^{-r}$ at particle-hole symmetry).

For a lattice DOS which is non-singular at the Fermi level, such a power-law behavior of the self-energy automatically causes the hybridization function to scale as $\Gamma(\w) \propto |\w|^r$ at low energies. In the context of single-impurity models, such a hybridization function defines a pseudogap Anderson model.


\subsection{Pseudogap Anderson and Kondo models}

The pseudogap Kondo problem, i.e., a Kondo impurity coupled to a fermionic bath with power-law DOS $\propto |\w|^r$, $r>0$, has been considered first in the context of $d$-wave superconductors.\cite{Withoff90} Subsequently, NRG has been used to thoroughly study the phase diagrams of both pseudogap Kondo and Anderson models\cite{Gonzalez98} and to determine spectral properties.\cite{Bulla97, Vojta01} Analytical renormalization-group techniques have led to an essentially complete understanding of the models' various quantum phase transitions.\cite{Fritz04a,Fritz04b} The case with divergent bath DOS, $r<0$, has been studied as well\cite{Mitchell13} (but is not relevant in the DMFT context).

For $r>0$ and in the presence of particle-hole symmetry, both Anderson and Kondo models display a quantum phase transition between a local-moment phase, dubbed LM, where the impurity is decoupled from the bath and hence unscreened, and a symmetric strong-coupling phase (dubbed SSC), where the impurity is partially screened and effectively behaves as an additional bath site. The impurity quantum phase transition between these two stable phases is governed by a symmetric critical fixed point (dubbed SCR). Importantly, SSC corresponds to a stable phase only for $r<1/2$, such that SCR exists only for bath exponents $0<r<1/2$. In contrast, for $r>1/2$ the impurity is never screened. We note that the situation with particle-hole asymmetry is richer, but of no relevance for the following discussion.

The low-energy behavior of the impurity spectral function across the phase diagram of the pseudogap Anderson and Kondo models is known both from numerical and analytical considerations. In short, for a bath DOS $\propto |\w|^r$, the impurity spectral function scales as $A_\imp(\w) \propto |\w|^r$ at the LM fixed point, whereas it behaves as $A_\imp(\w) \propto |\w|^{-r}$ at the SCR and SSC fixed points.\cite{Fritz04b, Vojta01, Bulla97}


\subsection{Self-consistent power-law solutions at zero temperature}
\label{sec:anapow}

Using Eq. \eqref{eq:BetheSC} we see that a DMFT spectral function scaling as $|\w|^r$ implies that both hybridization function and output spectrum of the Anderson impurity model must scale also as $|\w|^r$ -- this applies to any lattice DOS without low-energy singularity.
For the particle-hole-symmetric pseudogap Anderson model, an impurity spectrum $\propto |\w|^r$ can only be realized in the LM phase. We conclude that if a scale-invariant power-law solution of the DMFT equations exists, it must correspond to the LM phase of the impurity model. Hence, such criticality would \emph{not} be driven by an impurity critical point of the underlying impurity model, but instead by the DMFT self-consistency.

Within DMFT, the LM impurity phase corresponds to the insulating phase of the Hubbard model. This can have a power-law (gapless) spectrum only at the boundary to the metal, i.e., at the point when the Mott gap closes upon reduction of $U$. Given the DMFT phase diagram of the Hubbard model, we hence conjecture that the insulating solution of the DMFT equations, which exists for $U>U_{c1}$, becomes gapless as $U\searrow U_{c1}$ (consistent with earlier work \cite{Georges96}), with a power-law spectrum $A(\w)$. We recall that for $T\to 0$ the insulating solution is metastable\cite{unstable_note} in the entire range $U_{c1}<U<U_{c2}$.

From these qualitative considerations we cannot draw any conclusion concerning the spectral exponent $r$ of the conjectured power-law solution: The above exponent consideration is valid for any $0<r<\infty$, as long as the LM fixed point of the impurity model can be reached. This suggests that a concrete value of $r$ is selected via matching of power-law prefactors and hence might be non-universal. We will come back to this issue below.

For $U>U_{c1}$ the insulating solution develops a Mott gap $\Delta_U$ which can be expected to follow the critical scaling $\Delta_U \propto (U-U_{c1})^{\nu z}$. In the absence of spatial critical behavior, a separate definition of the correlation length exponent $\nu$ and the dynamical exponent $z$ is not meaningful, as only their product can be observed. We also note that, without an order parameter, the conventional critical exponents $\beta,\gamma,\delta$ cannot be defined. This leaves us with the exponents $\nu z$ and $r$; the latter being equivalent to an anomalous exponent of the single-particle spectrum.


\subsection{Critical insulator vs. critical metal}
\label{sec:critins}

The scale-invariant DMFT solution discussed above, with $\Sigma(\w) \propto |\w|^{-r}$ and $r>0$, corresponds to a critical insulator in the sense that the divergence of the self-energy suppresses the single-particle spectral density everywhere in momentum space. As a result, no signatures of a Fermi surface exist at low energies, and all excitations are fully incoherent due to strong inelastic scattering.

This can be contrasted to the scenario of critical Fermi surfaces at the Mott transition as put forward in Ref.~\onlinecite{Senthil08}: If the self-energy on the Fermi surface behaves as $\Sigma(\w) \propto |\w|^\alpha$ with $0<\alpha<1$ (but is non-critical elsewhere in momentum space), then the Fermi surface retains a power-law singularity; this corresponds to a critical metal (not unlike a Luttinger liquid in one space dimension). However, such a situation cannot be realized within DMFT, i.e., with a momentum-independent self-energy.


\subsection{Finite temperature and critical resistivity}
\label{sec:resfromspec}

Under the assumption that a power-law DMFT solution exists at $T=0$ at $U=U_{c1}^+$, we can discuss its implications at finite temperatures. According to standard quantum critical phenomenology, the power law is expected to be cutoff for frequencies smaller than $T$, such that $A(\omega, T) \propto T^{r} g(|\omega|/T)$ at $U_{c1}^+$, with $g(x)\propto\const$ for $x\ll 1$ and $g(x)\propto x^r$ for $x\gg 1$.

The single-particle spectrum fully determines the electrical conductivity due to the absence of vertex corrections. Inserting the above scaling form (or the corresponding scaling form for the self-energy) into Eq.~\eqref{eq:dmftsigma}, one immediately obtains for the dc conductivity at $U_{c1}^+$ the result $\sigma(T) \propto T^{2r}$. Hence, the power-law spectrum at $T=0$ implies a power-law divergence of the resistivity, consistent with the classification of the state as a critical insulator.


\section{Spectral power laws near $U_{c1}$: Numerics at $T\to 0$}
\label{sec:numzerot}

Motivated by the analysis of Sec.~\ref{sec:ana}, we now turn to a numerical solution of the DMFT equations. We study in detail the insulating solution close to $U_{c1}$, where it is metastable\cite{unstable_note} compared to the thermodynamically favorable metallic phase.
Within the DMFT self-consistency, the insulating solution near $U_{c1}$ and at low $T$ is attractive only in a very small region in the space of hybridization functions, thus it is difficult to access in an iterative scheme. This limits our accuracy to extract the asymptotic solution at $U_{c1}^+$, where the insulator becomes marginally unstable.
All calculations in this section were done at $T=10^{-8}$, which is much smaller than all other relevant energy scales and therefore reflects the physics of $T=0$.

\begin{figure}[t]
 \includegraphics[width=\columnwidth]{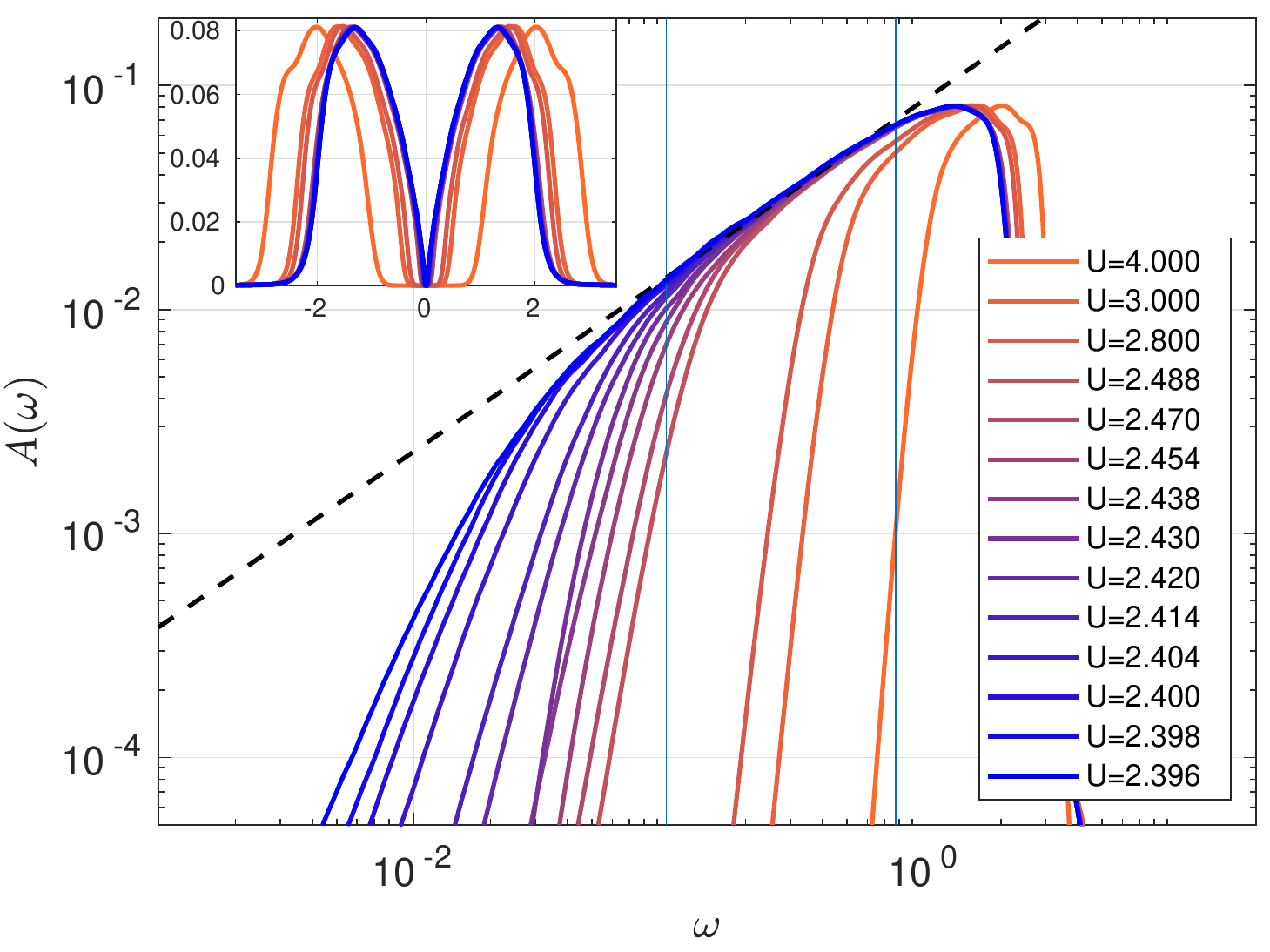}
 \caption{
 Hybridization function $A(\w)$ at $T=0$ for different $U>U_{c1}$. As the boundary of the coexistence region is approached, $A(\w)$ develops a power law with exponent $r_\hyb = 0.79(3)$ at intermediate frequencies (dashed line; fitting range indicated by vertical blue lines). The inset shows the same data on linear scale.
 }
 \label{fig:SF_T=0}
\end{figure}

\begin{figure}[b]
  \includegraphics[width=\columnwidth]{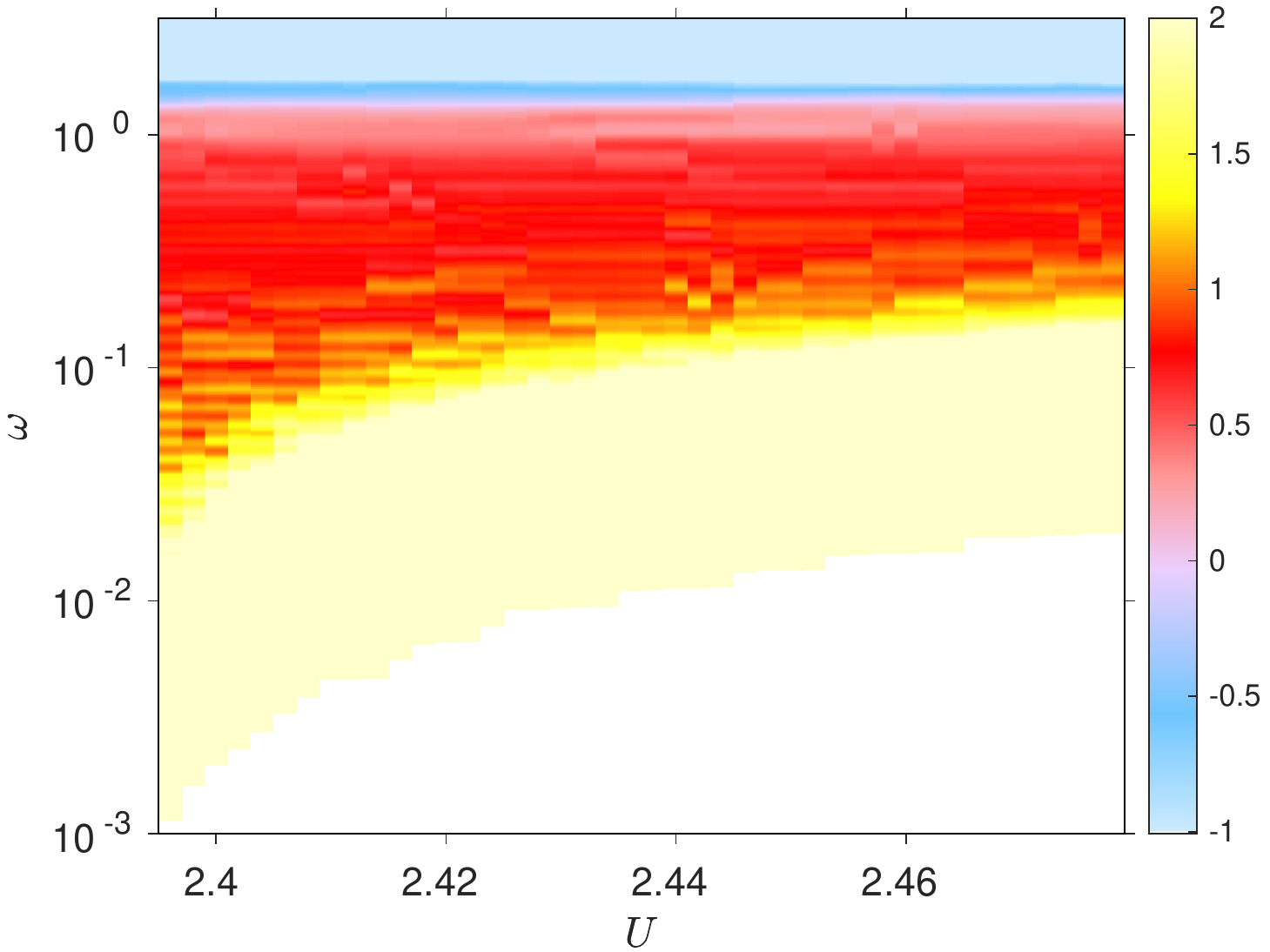}
  \caption{
  Effective exponent of the hybridization, $r = d \ln A(\omega) / d \ln \omega$, as function of $U$ and $\omega$ at $T=0$. Values larger than $r_{\max}=2$ and smaller than $r_{\min}=-1$ were set to $r_{\max}$ (light yellow) and $r_{\min}$ (light blue), respectively. The extent of the power-law region (red) grows as $U$ is decreased towards $U_{c1}$. Data have been cut off deep in the insulator where $A(\omega=2T)<10^{-8}$ (white).
  }
  \label{fig:heatmapT=0}
\end{figure}


\subsection{Hybridization function}
\label{sec:hybT=0}

DMFT results for the hybridization function $A(\w)$ are shown in Fig.~\ref{fig:SF_T=0}. We find that $A(\w)$ indeed displays a power-law regime over more than a decade in frequency, with exponent $r_\hyb\approx 0.8$, between the Hubbard bands as $U \searrow U_{c1}$. This power-law regime is clearly visible in the logarithmic derivative of $A(\w)$, Fig. \ref{fig:heatmapT=0}, as a region of constant color. We note that we were unable to obtain converged insulating solutions for $U<2.39$, i.e., very close to $U_{c1}$.

\begin{figure}[tb]
 \includegraphics[width=\columnwidth]{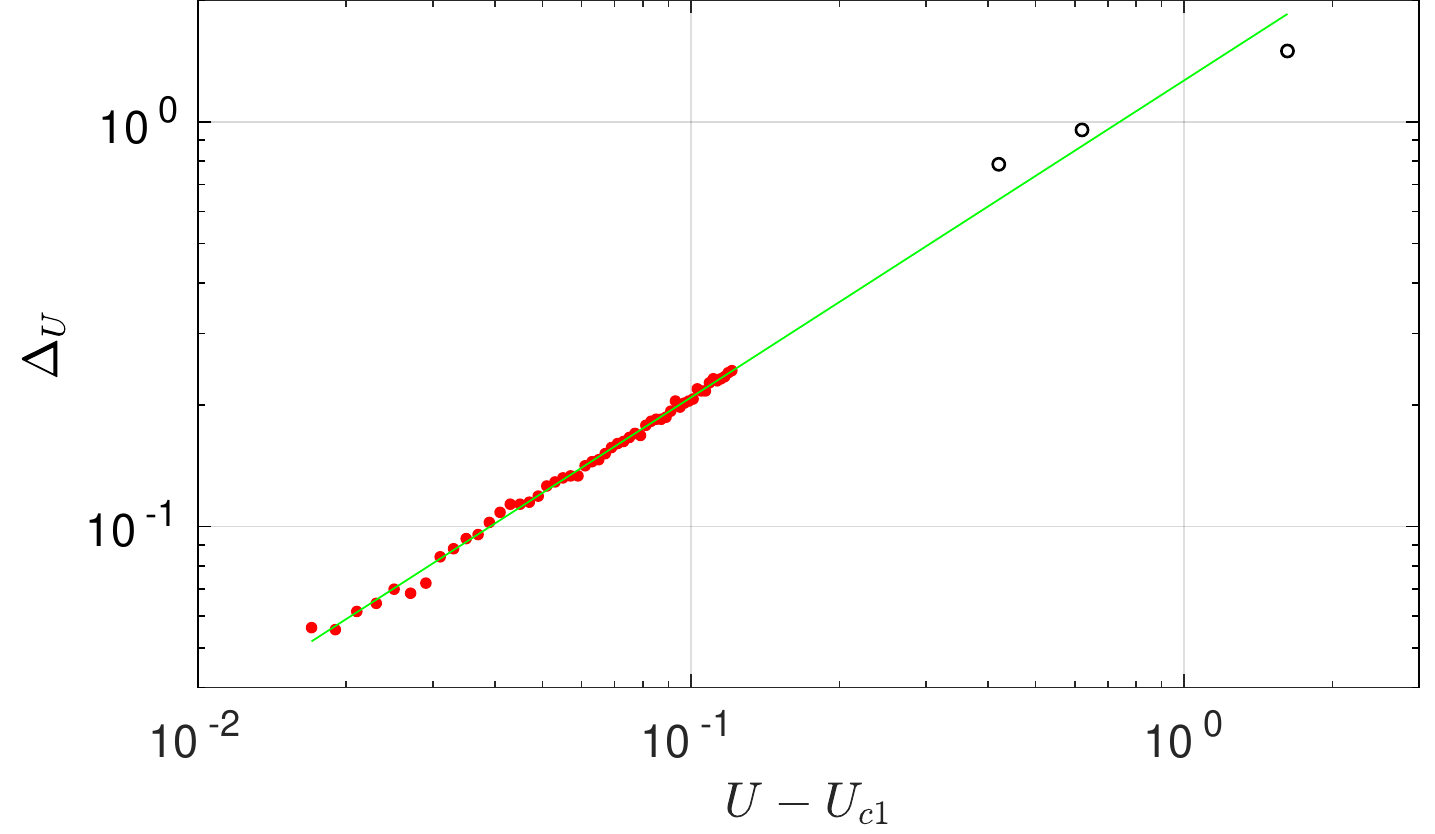}
 \caption{
 $T=0$ Mott gap $\Delta_U$ extracted from the spectra in Fig.~\ref{fig:SF_T=0} and shown as function of $(U-U_{c1})$. A non-linear fit (green) to $\Delta_U = c (U-U_{c1})^{\nu z}$ results in $c=1.3(2)$, $U_{c1}= 2.379(5)$ and $\nu z = 0.79(5)$. Only data points in the scaling regime (red) were included in the fit.
 }
 \label{fig:gapfit_T=0}
\end{figure}

For $U > U_{c1}$ the hybridization function develops a gap at small frequencies $|\w| \lesssim \Delta_U$. To extract this Mott gap $\Delta_U$ quantitatively, we extrapolate the (fitted) power law to $\w = 0$ and define $\Delta_U$ as the highest frequency where the relative deviation of $A(\w)$ from this power law is larger than\cite{gapcriterion_note} $15\%$.
The $U$ dependence of the extracted gap $\Delta_U$ is shown in Fig.~\ref{fig:gapfit_T=0}. Assuming a power-law dependence $\Delta_U \propto (U-U_{c1})^{\nu z}$ as expected near quantum criticality, we can obtain $U_{c1}$ and $\nu z$ by a non-linear fit including data for $2.39<U<2.5$, resulting in an exponent $z \nu \approx 0.79(5)$. This fit describes the data near $U_{c1}$ well, while deviations occur in the non-critical regime for larger $U$ where the spectrum does not develop an intermediate power law.
The fact that $\nu z$ is significantly smaller than unity shows that the low-energy behavior is beyond a rigid shift of Hubbard bands as a linear function of $(U-U_{c1})$.

\begin{figure}[bt]
 \includegraphics[width=\columnwidth]{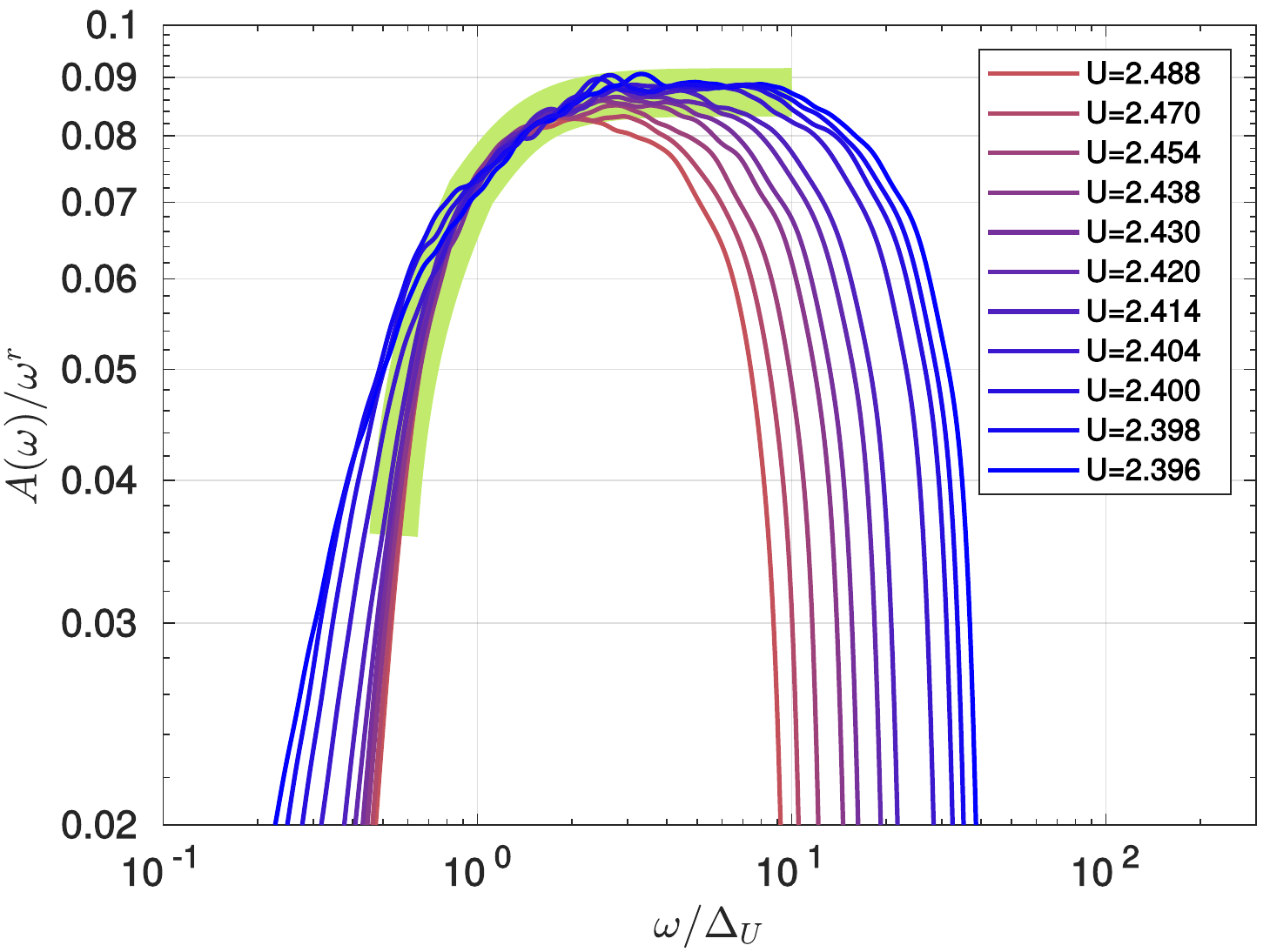}
 \caption{
   Hybridization function as shown in Fig. \ref{fig:SF_T=0}, but now plotted as $A(\omega)/ \omega^{r_\hyb}$ vs. $\omega / \Delta_U$, with $r_\hyb=0.784$ and $\Delta_U$ from Fig.~\ref{fig:gapfit_T=0}. While there is reasonable data collapse for $\omega / \Delta_U \lesssim 1$, there are systematic deviations from scaling both at large $U$ and for $U$ close to $U_{c1}$. The green shading is a guide to the eye highlighting the universal part of the data.
 }
\label{fig:SFcollapse_T=0}
\end{figure}

The hybridization functions in Fig.~\ref{fig:SF_T=0} suggest quantum criticality, and it is natural to test universal scaling. The expected scaling form  of the low-energy part of the spectrum near criticality at zero temperature is
\begin{equation}
A(\omega, U) \propto |\omega|^{r_{\hyb}} f_A\left(\frac{|\omega|}{\Delta_U}\right)
\label{eq:specscal}
\end{equation}
where $f_A(x)$ is a universal function.\cite{Sachdevbook} For large $x$ we must have $f_A\to\const$ so that for $\Delta_U \ll \w$ the spectral function follows the critical power law. The decay $f_A\to 0$ at small $x$  parameterizes how the gap opens away from criticality.
To test Eq.~\eqref{eq:specscal} numerically, we show in Fig.~\ref{fig:SFcollapse_T=0} the rescaled hybridization $A(\w)/\w^{r_\hyb}$ as function of $\omega/\Delta_U$ for $U$ close to $U_{c1}$. The data tend to collapse onto a universal curve, with systematic deviations occurring for large $\w$; this is expected as the corresponding data is outside the critical regime. However, deviations from scaling are also seen at small $\w$ and small $(U-U_{c1})$, indicating that the observed behavior is possibly not asymptotic.

\subsection{Self-energy}

\begin{figure}[bt]
 \includegraphics[width=\columnwidth]{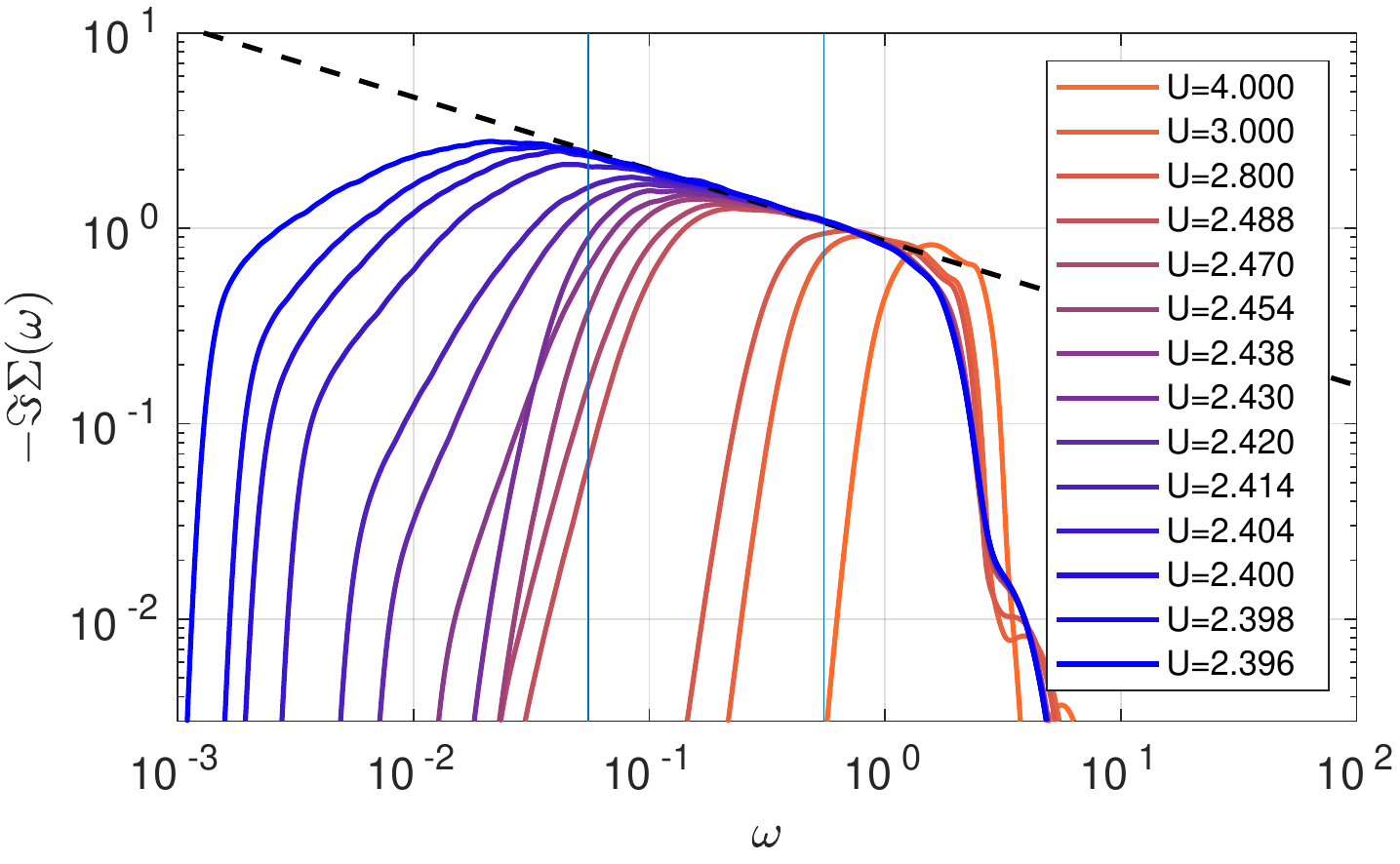}
 \caption{
 Imaginary part of the self-energy as function of frequency at different $U>U_{c1}$. Close to $U_{c1}$ a power law develops, but with an exponent $r_\Sigma = -0.37(4)$ deviating significantly from the expected value $-r_\hyb$ of an asymptotic power law  (dashed line; fitting range indicated by vertical blue lines).
 }
 \label{fig:Sigma_T=0}
\end{figure}

Further information can be gained by analyzing the DMFT self-energy, whose imaginary part is shown in Fig.~\ref{fig:Sigma_T=0}. Interestingly, $\Im \Sigma(\w)$ also develops a power law in the same frequency range as the hybridization, but with an exponent $r_\Sigma \approx -0.4$.

Clearly, power-law behavior reaching to lowest frequencies at $U_{c1}^+$ would require $r_\hyb=-r_\Sigma$, as dictated by DMFT self-consistency and Kramers-Kronig relations. We conclude that the power laws in Figs.~\ref{fig:SF_T=0} and \ref{fig:Sigma_T=0} are not asymptotic, but represent crossover behavior. This naturally explains the deviations from scaling close to $U_{c1}$ which are visible in Fig.~\ref{fig:SFcollapse_T=0}.

Constructing an ``asymptotic'' self-energy from Fig.~\ref{fig:Sigma_T=0} by extrapolating the $|\w|^{-0.4}$ behavior all the way down to $\w=0$, yields, according to the Dyson equation, a single-particle spectrum scaling as $|\w|^{0.4}$ for $\w\lesssim 10^{-3}$, i.e., for frequencies far below the smallest gap reached in Fig.~\ref{fig:SF_T=0}. This implies that simulations much closer to $U_{c1}$ (where we have been unable to reach convergence) would be required to see such an asymptotic power law.


\subsection{Universality}

To study the universality of these phenomena, similar calculations were performed with other bare DOS $\rho(\epsilon)$, with details shown in Appendix~\ref{app:otherdos}. For all tested DOS, we obtained crossover power laws in both $A(\w)$ and $\Sigma(\w)$ as for semicircular DOS. The fitted exponents were in the range $r_\hyb = 0.77 \ldots 0.87$ and  $-r_{\Sigma}= 0.36 \ldots 0.43$. We note that these exponents have error bars of order 0.05 due to the limited fitting range.

At this point we cannot decide whether the spectral exponents are universal or not. Given that we are analyzing a crossover phenomenon, universality in the strict sense is unlikely to be expected; we also recall the corresponding remarks in Sec.~\ref{sec:anapow}. However, even if not universal, the numerical exponent values turn out to be remarkably robust.


\subsection{$T=0$ summary}

We have shown that at $T=0$ the insulating solution of the DMFT equation near $U_{c1}$ displays a power law in an intermediate range of frequencies, $0.03 \lesssim \w \lesssim 0.8$. It is a crossover power law, since the exponents of the hybridization and the self-energy are not consistent with an asymptotic solution.
For smaller gaps, $\Delta_U \lesssim 10^{-3}$, we conjecture that the hybridization function shows a crossover to the exponent $r \approx 0.4$, such that an asymptotic power law with $r_\hyb=-r_\Sigma$, as postulated in Sec.~\ref{sec:ana}, emerges. However, we were unable to access this regime numerically due to the marginal instability of the insulating solution at $U_{c1}^+$.

Given that the discovered crossover power law extends to rather high frequencies nearly up to the Hubbard band, it can be expected to influence the physics at elevated temperatures. In contrast, the conjectured asymptotic power law is limited to such a small range of $U$ and $\w$ that it will be unobservable except at very low $T$ and hence is irrelevant to the crossover physics above the critical endpoint.


\begin{figure*}
 \includegraphics[width=\textwidth]{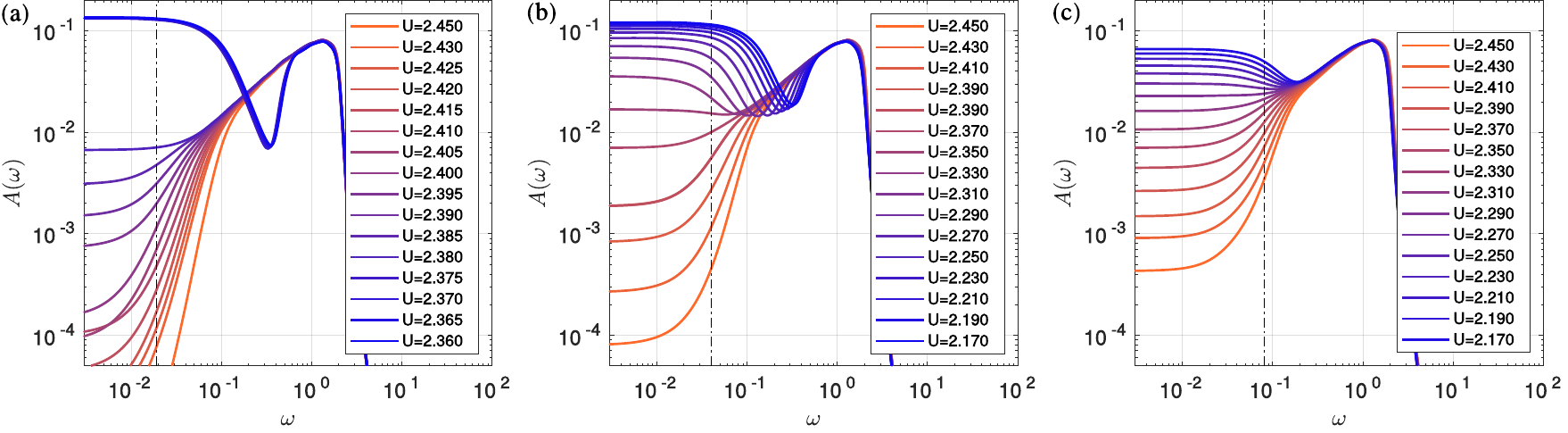}
 \caption{
  Hybridization function for (a) $T=0.0192 \approx 0.6 \, T_c$, (b) $T=0.04 \approx 1.3 \, T_c$ and (c) $T=0.08 \approx 2.7 \, T_c$.
 In (a) a power law is visible at intermediate frequencies for larger $U$, i.e., for insulating-like solutions, and a first-order transition to the metal occurs as $U$ is lowered.
 In (b) and (c) a power law is visible at intermediate frequencies in the crossover between metallic and insulating phase; this crossover gets washed out as temperature is increased.
 The vertical lines mark the temperatures.
 }
\label{fig:SF_TgtrTc}
\end{figure*}

\begin{figure*}
 \includegraphics[width=\columnwidth]{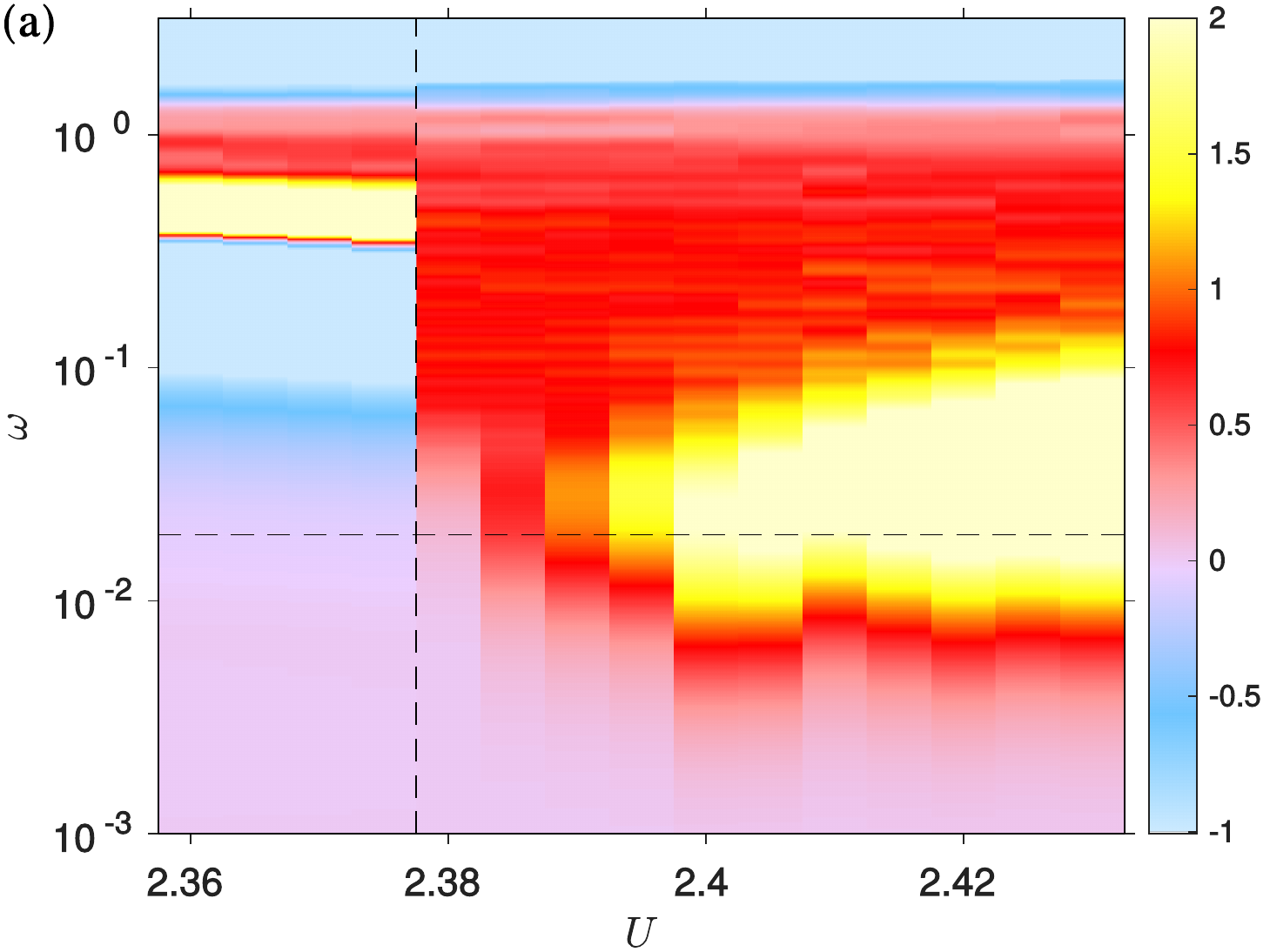}
 \includegraphics[width=\columnwidth]{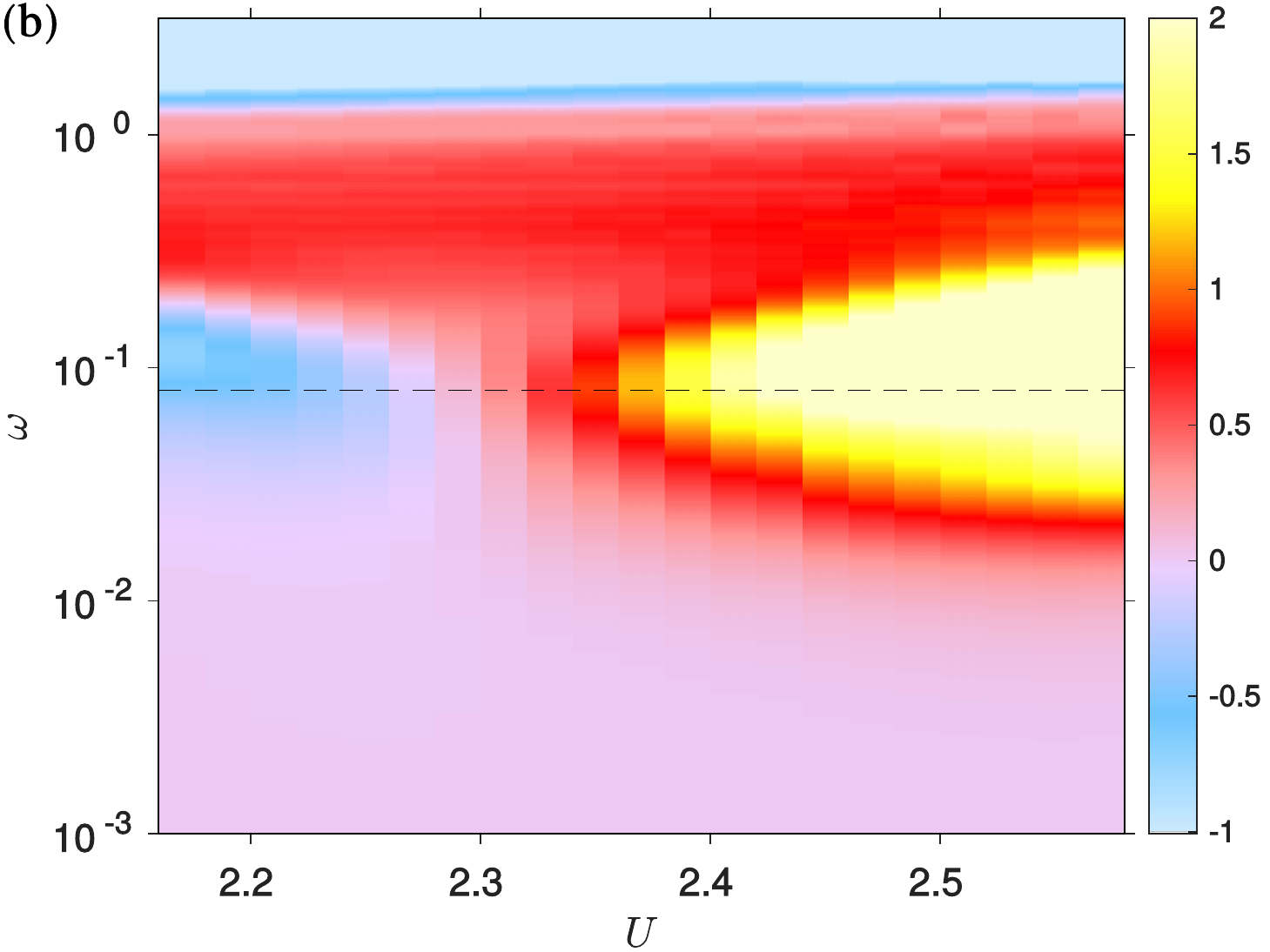}
 \caption{
 Logarithmic derivative $d \ln A(\omega) / d \ln \omega$ as function of $U$ and $\omega$ at (a) $T=0.0192 \approx 0.6 \, T_c$ and (b) $T=0.08 \approx 2.7 \, T_c$.
The horizontal dashed line marks the temperature, the vertical dashed line in (a) the boundary of the coexistence region $U_{c1}(T)$.
The power-law region (red) reaches its maximum extent close to $U_{c1}(T)$.
Note that in (b) the exponent in the power law region is slightly smaller than in (a).
}
\label{fig:heatmapT}
\end{figure*}

\section{Finite-temperature crossovers and quantum critical scaling}
\label{sec:numfinitet}

We now turn to a study of the Hubbard-model DMFT solution at finite temperatures. Our main goal is to understand the origin of the apparent quantum critical scaling of the resistivity in the regime above the finite-temperature endpoint, located at $T_c \approx 0.03$. To this end, we will track the physical properties from the low-temperature limit all the way to temperatures far above $T_c$.

\subsection{Hybridization function}
\label{sec:hybT}

We start by studying the single-particle spectrum at finite temperatures. For $T<T_c$ our focus is on the insulating solution near $U_{c1}$ where we expect the Mott gap to be thermally filled. For $T>T_c$ we study a range of $U$ values around $U_{c1}$ to cover the metal-insulator crossover.

Single-particle spectral functions for different $U$ at selected temperatures are shown in Fig.~\ref{fig:SF_TgtrTc}. The power law in $A(\w)$, with exponents $r_\hyb$ in the range $0.7\ldots0.8$, survives at finite temperatures, roughly in the frequency range $ \min(\Delta_U, T) \lesssim \omega \lesssim \min(U/2,W)$. For $T<\Delta_U$ the gap opening is followed by a thermal filling of the gap as $\w$ decreases; for $T\gtrsim\Delta_U$ a gap effectively never opens; in both cases $A(\w)\to\const$ for $\w\ll T$.
The crossover power law is nicely seen in the logarithmic derivative of $A(\w)$, Fig.~\ref{fig:heatmapT}. Given the thermal broadening on the scale $T$, a good indicator is the effective exponent at frequencies $\omega=2 T$, as shown in the phase diagram in Fig.~\ref{fig:fancyphasediagram}.
In addition, we note that for temperatures slightly above $T_c$, the crossover between metallic and insulating solutions is very sharp, Fig.~\ref{fig:SF_TgtrTc}(b), and the solutions on the metallic side are rather distinct from the insulating ones. At higher temperatures this difference is washed out, compare Fig.~\ref{fig:SF_TgtrTc}(c).

The relatively narrow frequency interval of the crossover power law makes the extraction of the exponent difficult and limits the accuracy of the comparison to the $T=0$ value. Nevertheless we see a trend that the exponent $r_\hyb$ decreases slightly as the temperature increases, from $r_\hyb \approx 0.8$ at $T=0$ to $r_\hyb \approx 0.7$ at $T=0.08$.
Again, we are unable to decide whether this implies non-universality of the exponents, as explained in Sec.~\ref{sec:ana}, or whether this is simply a result of corrections to scaling and fitting limitations.


\subsection{Quantum Widom lines}

The Widom line is commonly used to characterize phenomena in the vicinity of the endpoint of first-order phase transitions; it appeared first in the context of the liquid-gas transition where it marks the (supercritical) crossover between more liquid-like and more gas-like states. \cite{Widom69}
Refs.~\onlinecite{Dobro11,Dobro13} argued that a (quantum) Widom line $U^\ast(T)$ is needed as a reference line for the quantum critical resistivity scaling performed above the first-order Mott critical endpoint, Eq.~\eqref{eq:dobroscale}.

Physically, the Mott quantum Widom line should mark the crossover between the metallic and insulating regimes at finite $T$. Ref.~\onlinecite{Dobro11} introduced the Widom line as the point of maximum instability and correspondingly employed a criterion based on the convergence rate of the DMFT iteration. This led to a Widom line which emanates from the critical endpoint and shows non-monotonic behavior of $U^\ast(T)$, i.e., backbending towards large $U$ at elevated $T\gtrsim 3 \, T_c$. Other criteria for the choice of Widom line were discussed as well, which resulted -- in the temperature range $2 \, T_c \ldots 4 \, T_c$ -- in nearby lines, most of which displayed backbending.
In Ref.~\onlinecite{Dobro17} it was argued that the Widom line is linked to the thermal filling of the Mott gap, i.e., an instability of the insulator, to be contrasted to a thermal instability of the metal, which may be defined via the coherence temperature of the Fermi liquid or the temperature where resilient quasiparticles are destroyed.\cite{georges13}

While these arguments are plausible, a unique definition of a Widom line -- marking a crossover, not a transition -- cannot be expected. For instance, the authors Ref.~\onlinecite{McKenzie17} employed a vertical reference line, $U^\ast=\const$, in their scaling analysis of the dynamical spin susceptibility above the Mott endpoint. We note that $U^\ast=\const$ corresponds to standard quantum critical scaling.\cite{Sachdevbook}
On a pragmatic level, one can ask which reference line delivers the best approximate scaling collapse,\cite{scale_note} and the answer may depend on the observable considered, not the least due to background contributions and subleading corrections.
In our scaling analysis of the resistivity described below, we have tested several choices of Widom lines. Interestingly, we found a very convincing scaling collapse for a vertical reference line $U^\ast(T)=\const$ located at the zero-temperature instability point of the insulator. Backbent Widom lines also led to a reasonable scaling collapse, but at the expense of having a $\rho_c(T)$ with a behavior different above and below $T_c$, see next subsection.


\subsection{Resistivity and scaling}

\begin{figure}[tb]
 \includegraphics[width=\columnwidth]{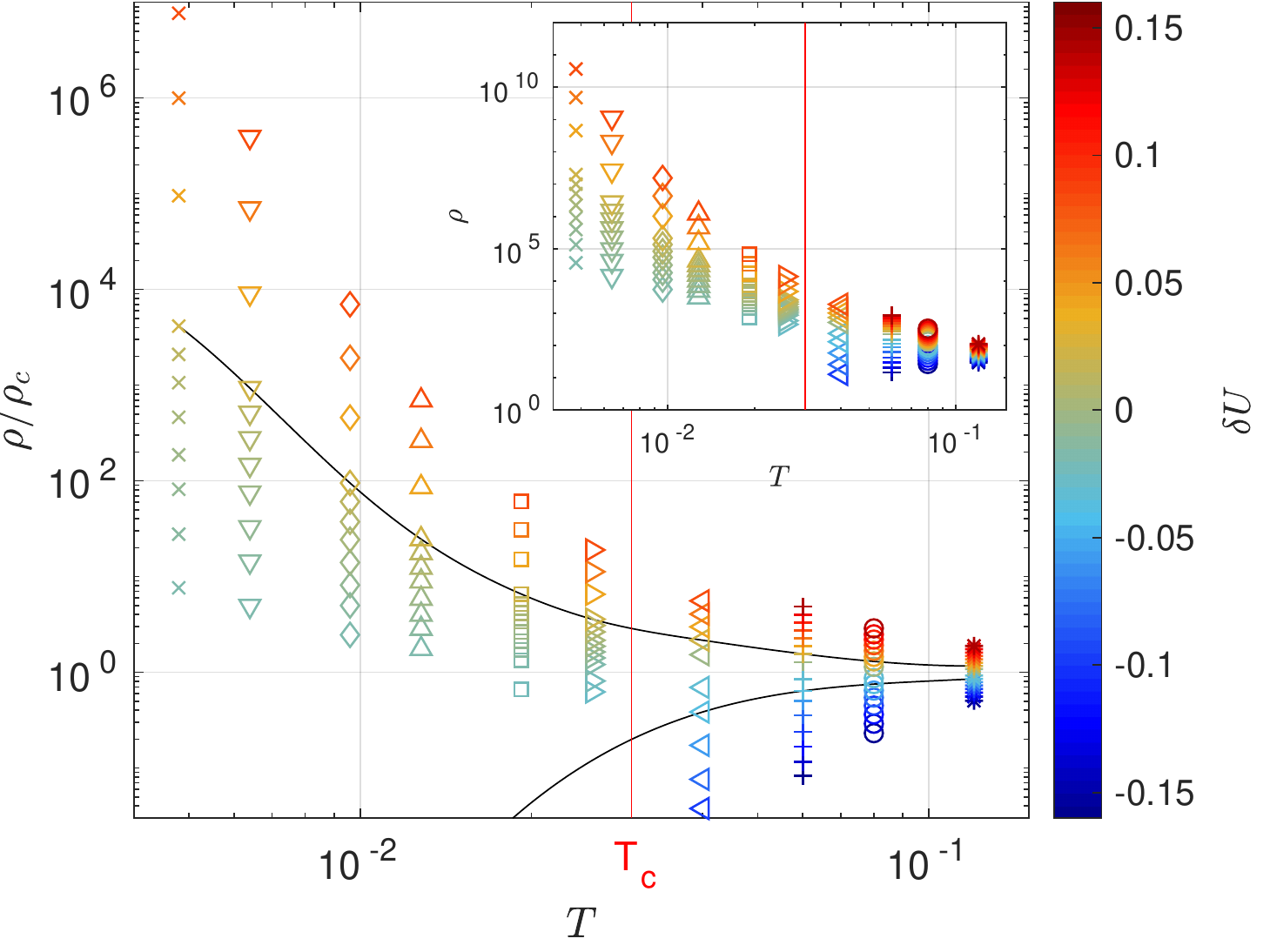}
 \caption{
 Resistivity on lines parallel to the Widom line with different $\delta U = U-U^\ast(T)$ as a function of temperature. The resistivity is normalized to $\rho_c(T)$, the resistivity on the Widom line $U^\ast(T)= 2.386$. The black curves mark the reference curve onto which all data points are collapsed; it is obtained by a spline interpolation of the data for $\delta U=\pm 0.04$.
 The different colors correspond to different values of $\delta U$, as shown in the colorbar. The inset shows the unscaled data for $\rho(\delta U,T)$.
 }
  \label{fig:resistivity}
\end{figure}

\begin{figure}[tb]
 \includegraphics[width=\columnwidth]{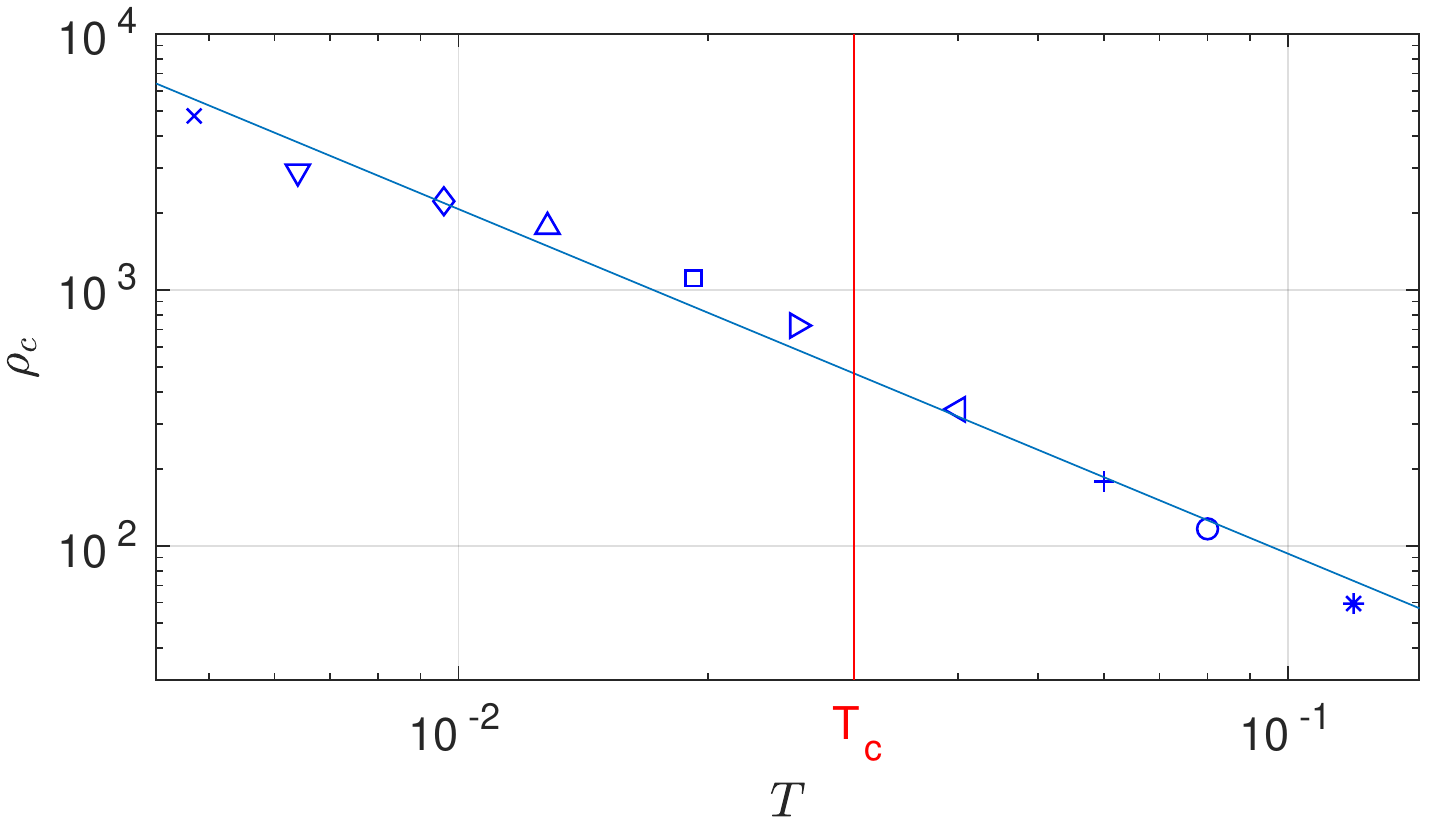}
 \caption{
 Value of the resistivity $\rho_c(T)$ on the Widom line $U^\ast(T) = 2.386$, used to normalize the resistivity in Fig. \ref{fig:resistivity}. It shows a power-law dependence on temperature both above and below the critical point, with exponent $-1.3(1)$.
 }
 \label{fig:scaling_rhoc}
\end{figure}

Results for $\rho(U,T)/\rho_c(T)$ are shown in Fig.~\ref{fig:resistivity} where a simple vertical Widom line located very close to $U_{c1}(T\!=\!0)$ has been used, $U^\ast(T) = 2.386$. Data\cite{benchmarkrho_note} are shown both above and below $T_c$, for the latter on the insulating side only.
The resistivity on the Widom line, $\rho_c(T)$ is displayed in Fig.~\ref{fig:scaling_rhoc}. It follows a power law to a good accuracy, $\rho_c \sim T^\alpha$ with $\alpha= -1.3(1)$. The low-temperature divergence naturally fits with the expected insulating behavior as $T\to 0$ at $U\searrow U_{c1}$, see Sec.~\ref{sec:critins}; its exponent is consistent with the relation $\alpha=-2r$ derived in Sec.~\ref{sec:resfromspec} recalling that $r_{\rm hyb}\approx 0.7$.

\begin{figure}[tb]
 \includegraphics[width=\columnwidth]{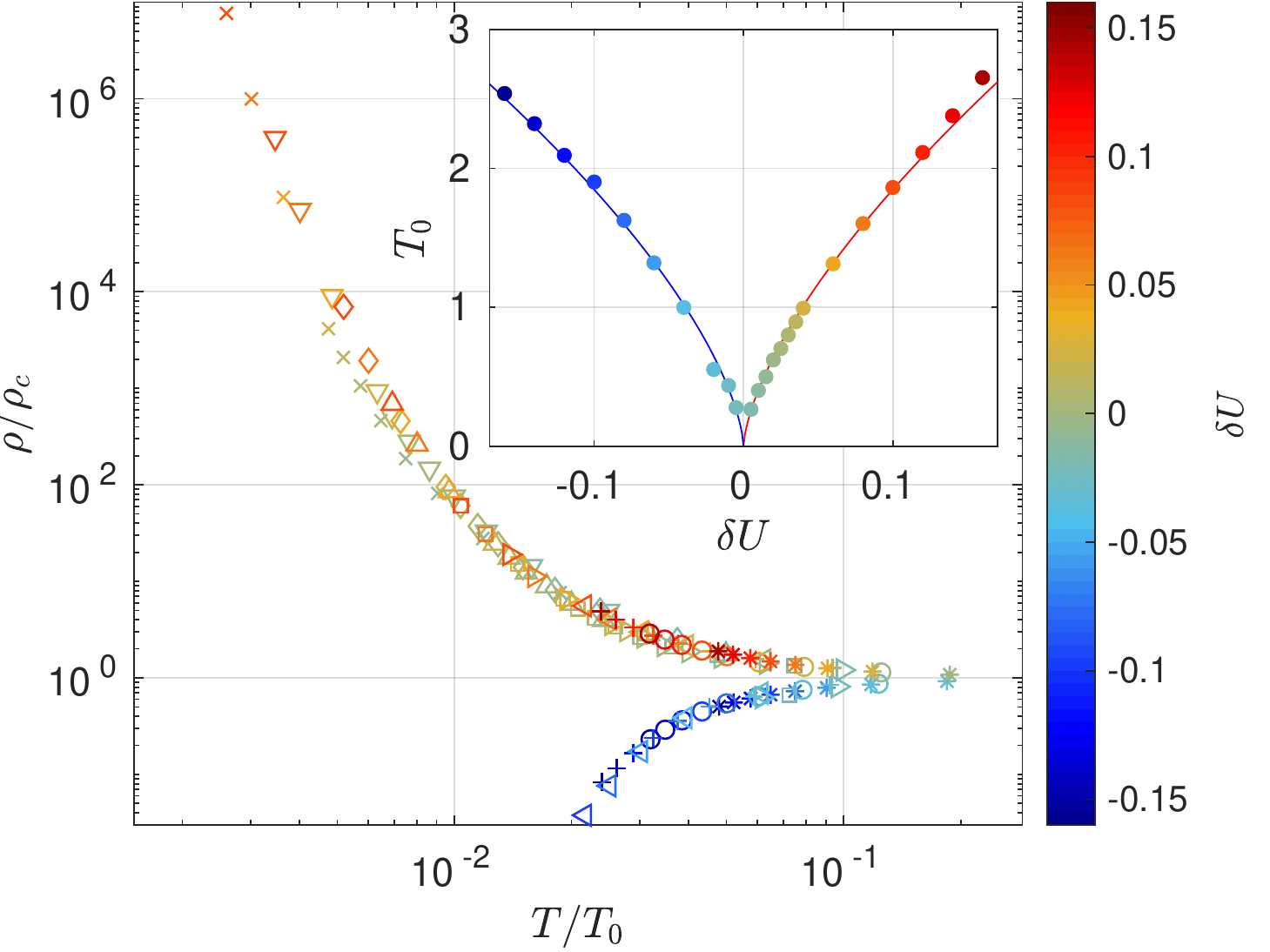}
 \caption{
 Collapse of data points if plotted as $\rho/\rho_c$ vs $T/T_0$. The color code and markers are the same as in Fig. \ref{fig:resistivity}. The scaling collapse is convincing, with deviations visible at low $T$ and large $\delta U$ where the resistivity data are numerically less reliable.
 Inset: Power-law fits to the rescaling factors $T_0$ give $\nu z_+ =0.66(2)$ for $\delta U >0$ and $\nu z_- =0.65(4)$ for $\delta U <0$.
}
 \label{fig:scaling_collapse}
\end{figure}

We now turn to a scaling analysis of the resistivity according to Eq.~\eqref{eq:dobroscale}, as described in detail in Ref.~\onlinecite{Dobro11}. To this end, we choose two values of $\delta U = U-U^\ast(T)$, namely $\delta U=\pm0.04$, which we use to define reference curves of $\rho/\rho_c$ and rescale the $\rho$ data at other $\delta U$ to these reference curves. The corresponding temperature rescaling factors $T_0$ (averaged over $T$ for each $\delta U$) are shown in the inset of Fig.~\ref{fig:scaling_collapse}. A power-law fit to $T_0 \propto |\delta U|^{\nu z}$ yields $\nu z_+ = \nu z_- = 0.66\pm0.1$, with the uncertainty estimated using different reference curves and scaling procedures. These values are roughly consistent with the exponent $\nu z$ reported in Sec.~\ref{sec:hybT=0} for the opening of the Mott gap at $T\to 0$; minor differences are likely rooted in the fact that we are not dealing with asymptotic critical behavior, but with a crossover phenomenon.

The rescaled resistivities, i.e., $\rho(\delta U, T)/\rho_c(T)$ as function of $T/T_0(\delta U)$, are plotted in Fig.~\ref{fig:scaling_collapse}. The scaling is found to be fulfilled to very good accuracy; the quality of the resulting scaling collapse is not changed visibly when $\nu z$ is changed by $\pm 0.05$ or less. Data points at high temperatures, i.e., in the scaling regime introduced by Ref.~\onlinecite{Dobro11}, and data points at low temperatures in the metastable insulating phase collapse onto the \emph{same} universal curve.

This consistent resistivity scaling both above and below $T_c$, Fig.~\ref{fig:scaling_collapse}, is the most important result of our work. It demonstrates that the behavior in the high-$T$ regime above the critical endpoint can be seamlessly connected to the low-$T$ metastable insulating regime. Together with the analysis in Secs.~\ref{sec:hybT=0} and \ref{sec:hybT} which demonstrate that the spectral power law found at low $T$ also continues to elevated $T$, our results strongly suggest that the low-$T$ approximately scale-invariant solution near $U_{c1}$ is responsible, and thus drives, the Mott quantum critical transport.

The exponent $\nu z \approx 0.7$ is somewhat larger than the values reported from the resistivity scaling in Refs.~\onlinecite{Dobro11,Dobro13,Kanoda15}; this is due to the different choice of $U^\ast(T)$.
We note that for other Widom lines defined in the same region of the phase diagram, with or without back-bending, the scaling collapse can be of a similar quality as in Fig.~\ref{fig:scaling_collapse}, with exponents $\nu z_\pm$ in the range  $0.5 \ldots 0.8$ (not shown), similar to that reported in Refs.~\onlinecite{Dobro11, Dobro13} but also including data for $T<T_c$. However, the temperature dependence $\rho_c(T)$ along the Widom line will in general not be a simple power law, because $\rho_c$ must diverge as $T\to 0$, but its behavior above $T_c$ strongly depends on the choice of Widom line. For example, $\rho_c(T)$ on the Widom lines discussed in Ref.~\onlinecite{Dobro13} exhibits only a weak temperature variation in the small interval $[2 \, T_c, 4 \, T_c]$ considered there.


\section{Conclusions and outlook}
\label{sec:concl}

With the goal to understand the apparent quantum critical scaling of the resistivity above the critical endpoint of the first-order Mott transition, we have studied in detail spectral properties of the DMFT solution to the one-band Hubbard model at half-filling.
We have established that DMFT admits solutions with low-energy power-law spectra corresponding to a metastable insulator in the limit $U\searrow U_{c1}$. Our numerics have uncovered that this asymptotic regime (which we have not been able to access in detail) is narrow in both frequency and temperature, rendering it irrelevant for the observed resistivity scaling.
However, we have discovered a wide crossover regime where the spectrum displays a distinct power law: This crossover regime extends to temperatures above $T_c$ and can be connected to the scaling.
We have studied the resistivity for $T<T_c$ and $U>U_{c1}$ as well as above the critical endpoint, and showed that it displays common scaling behavior. We conclude that the apparent quantum critical resistivity scaling is rooted in the wide power-law crossover regime characteristic of the insulating phase near the Mott transition.

Let us turn to broader implications of the present results and those of Refs.~\onlinecite{Dobro11,Dobro13,Kanoda15,Dobro17}:
First, the fact that DMFT theory and experiments agree so well implies that the dominant correlation physics in the regime above the critical endpoint is local in space. It is clear that, in experiments, non-local effects must set in at sufficiently low $T$, e.g. by producing magnetic order or spin-liquid-like correlations between local moments, but the corresponding scale is significantly smaller than $T_c$.
Second, we have argued that the local physics of the Mott transition above this scale can be understood as an instance of approximate local quantum criticality, driven by strong inelastic scattering in a fully incoherent regime, and DMFT has enabled us to make the connection to low temperatures explicit.
Third, the non-local physics at low $T$ will be lattice-dependent, and this calls both for experiments on tunable Mott compounds on lattices other than triangular and for numerics beyond DMFT, e.g. using cluster DMFT, on different lattices. It has to be kept in mind that frustration is needed to suppress magnetic order at elevated temperatures.
Finally, a thorough investigation away from half filling is required: Mottness can be destroyed by doping, and previous DMFT studies\cite{Dobro15} indicated that a finite-temperature resistivity scaling regime also emerges in this case. It is crucial to clarify whether this scaling regime can be connected to the same or a different mechanism of local quantum criticality as the half-filled case studied here, and to understand its relation to the physics of doped cuprates and other carrier-doped oxides.


\acknowledgments

We thank A. Weichselbaum for contributing major parts of the code used in this work.
We acknowledge helpful discussions with J. von Delft, F. Anders, A. Weichselbaum, and in particular V. Dobrosavljevi\'{c} as well as numerics support by R. Peters in an early stage of this project.
HE and MV acknowledge financial support from the DFG through SFB 1143 (project-id 247310070) and the W\"urzburg-Dresden Cluster of Excellence on Complexity and Topology in Quantum Matter -- \textit{ct.qmat} (EXC 2147, project-id 39085490).
SSBL acknowledges support from the DFG through the Munich Center for Quantum Science and Technology (EXC 2111, project-id 390814868) and from Grant No.~LE 3883/2-1.
The calculations were performed at the computer cluster of the Arnold Sommerfeld Center for Theoretical Physics.


\appendix

\section{Universality of exponents}
\label{app:otherdos}

\begin{figure}
 \includegraphics[width=\columnwidth]{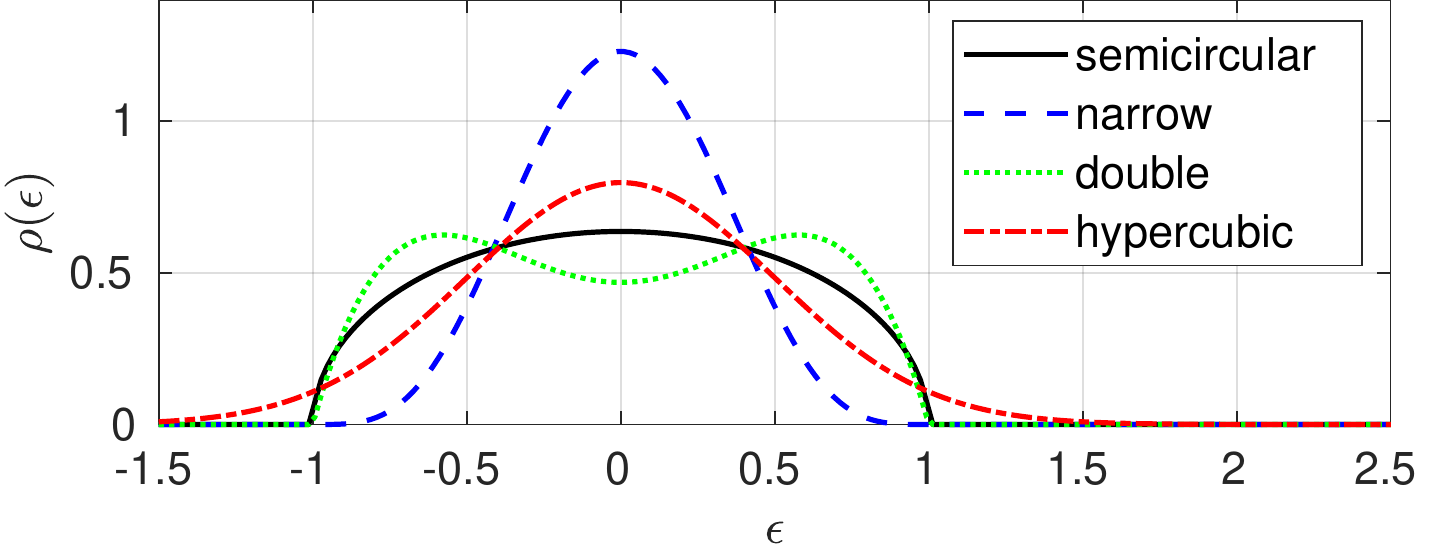}
 \caption{
 Different densities of states used to test the universality of the spectral exponents, for details see text.
  }
 \label{fig:otherDOS}
\end{figure}

\begin{figure}
 \includegraphics[width=\columnwidth]{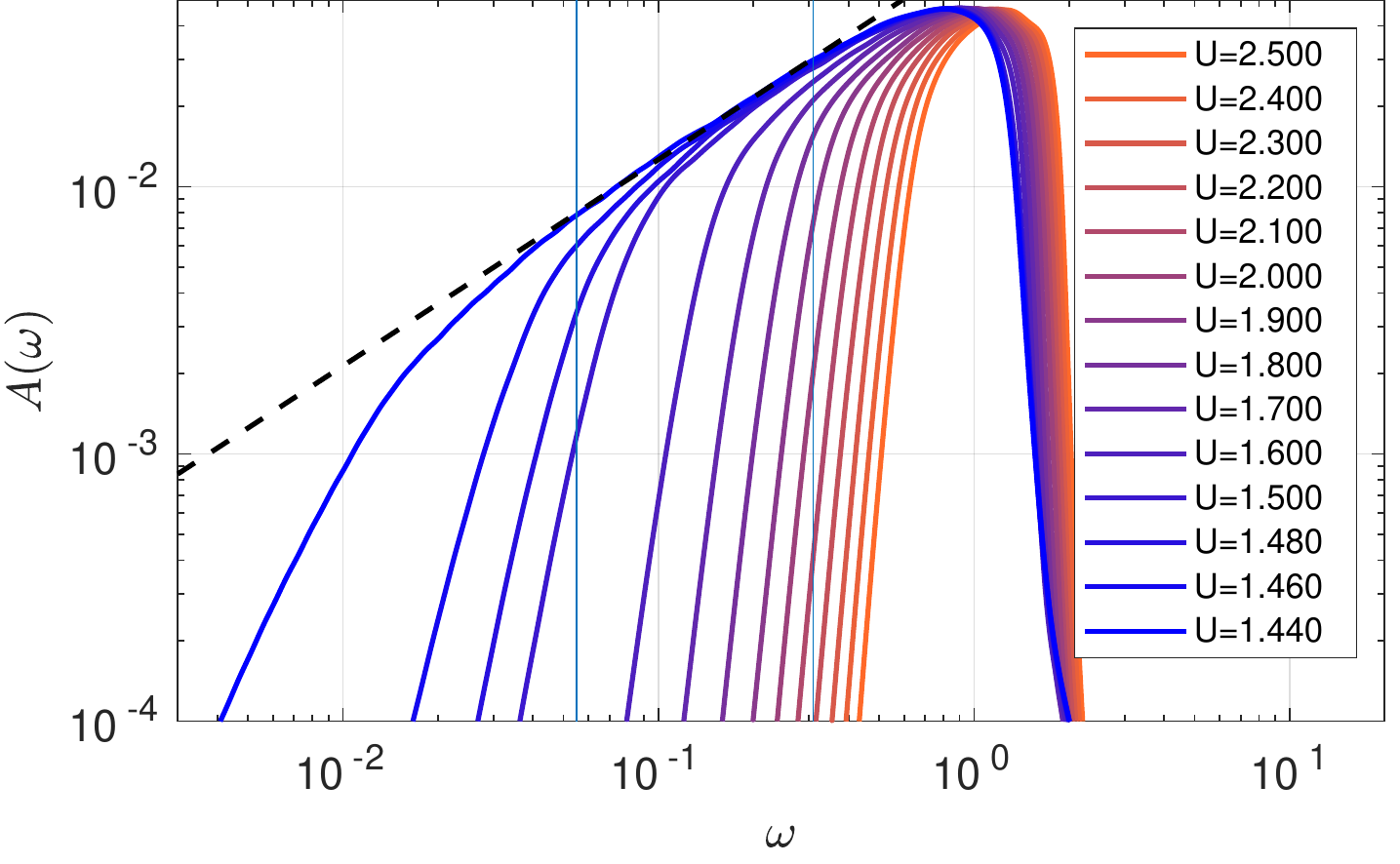}
 \caption{
 Hybridization at $T=0$ as function of frequency at different $U>U_{c1}$ for the DOS $\rho_{\rm narrow}(\epsilon)$ (see text). The behavior is qualitatively similar to the semicircular case, but the exponent of the crossover power law is slightly different. The dashed line shows a power-law fit with $r_{\rm hyb} = -0.77(3)$ (fitting range indicated by vertical blue lines).
 }
 \label{fig:SF_T=0_narrowpeak}
\end{figure}

\begin{figure}[b]
 \includegraphics[width=\columnwidth]{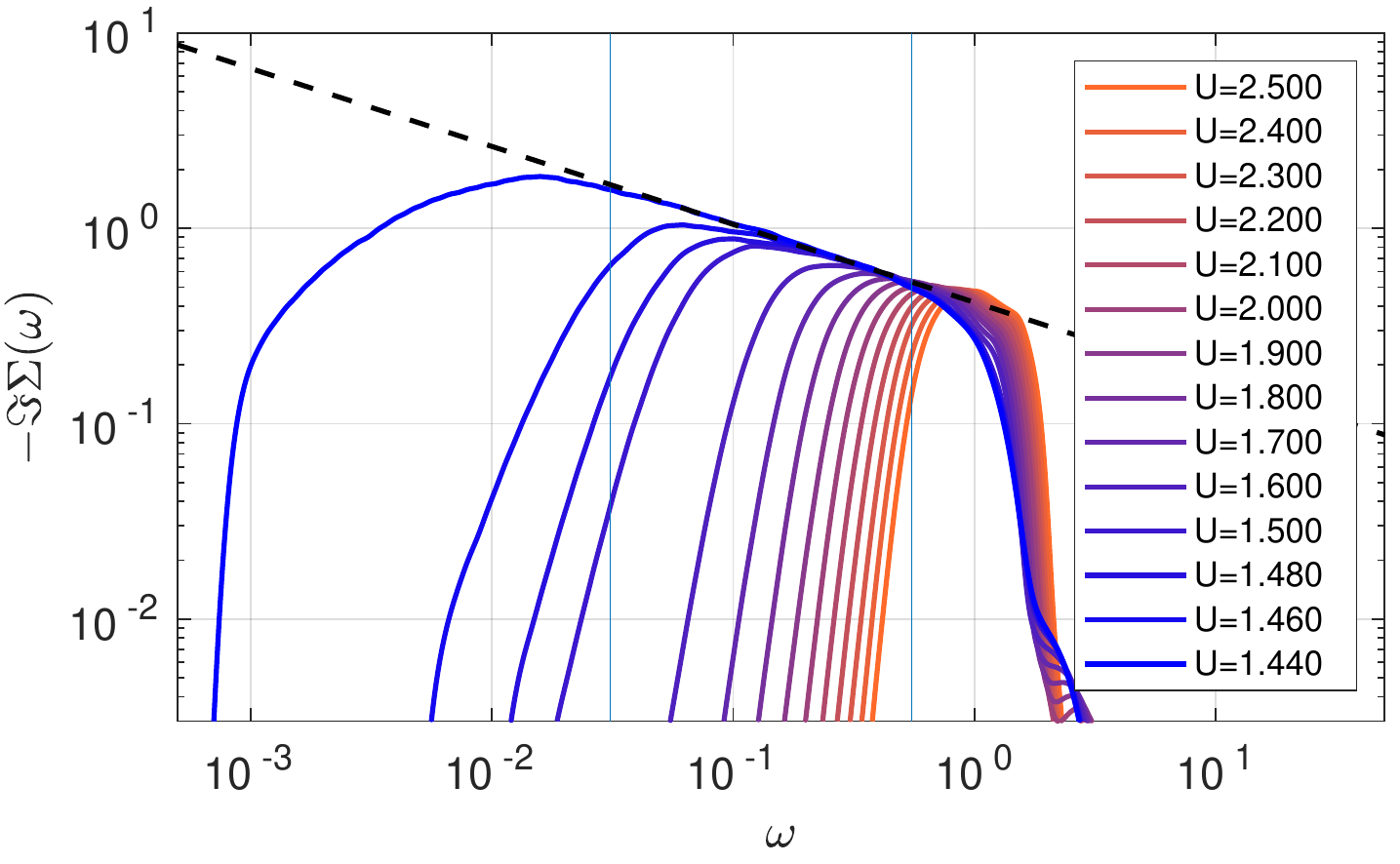}
 \caption{
 Imaginary part of the self-energy as function of frequency at different $U>U_{c1}$ for the DOS $\rho_{\rm narrow}(\epsilon)$. The dashed line shows a power-law fit with $r_\Sigma = -0.40(2)$ (fitting range indicated by vertical blue lines).
 }
 \label{fig:Sigma_T=0_narrowpeak}
\end{figure}

In addition to the semicircular Bethe-lattice DOS
\begin{equation}
\rho(\epsilon) = 2 \sqrt{W^2-\epsilon^2}/\pi W^2
\end{equation}
we have also studied other DOS, Fig.~\ref{fig:otherDOS}, in order to check for universality of the behavior, in particular of the power-law exponents. We have employed
\begin{align}
\rho_{\rm narrow}(\epsilon) &= \frac{315}{256} \, \Theta(W-|\epsilon|) \left(1- \frac{\epsilon^2}{W^2}\right)^4,
\end{align}
\begin{align}
\rho_{\rm double}(\epsilon) &= \frac{15}{16} \, \Theta(W-|\epsilon|) \left(\frac{1}{2} + \frac{\epsilon^2}{W^2} -\frac{3}{2}  \frac{\epsilon^4}{W^4}\right),
\end{align}
\begin{align}
\rho_{\rm hc}(\epsilon) &= \frac{2}{W \sqrt{2 \pi}} \exp(- 2 \epsilon^2/W^2).
\end{align}
Here both $\rho_{\rm narrow}(\epsilon)$ and $\rho_{\rm double}(\epsilon)$ have the same support as the semicircular DOS $[-W,W]$, but $\rho_{\rm narrow}(\epsilon)$ has more weight close to the Fermi level while in $\rho_{\rm double}(\epsilon)$ weight is shifted towards the band edges. Finally $\rho_{\rm hc}(\epsilon)$ corresponds to a hypercubic lattice in infinite dimensions. Parenthetically, we note that Hubbard model on the Bethe lattice with second-neighbor hopping has been studied using DMFT in Ref.~\onlinecite{Peters09}, and DOS with van-Hove singularities have been studied in Ref. \onlinecite{Pruschke09}.

Numerical results for the hybridization and self-energy obtained for $\rho_{\rm narrow}(\epsilon)$ are shown in Figs.~\ref{fig:SF_T=0_narrowpeak} and \ref{fig:Sigma_T=0_narrowpeak}, as before focusing on $T=0$ and $U\gtrsim U_{c1}$. The qualitative behavior is similar to that of for the semicircular DOS, with crossover power laws in both $A(\w)$ and $\Sigma(\w)$. We obtain $U_{c1}=1.43$, $r_{\hyb}=0.77(3)$ and  $r_{\Sigma}=-0.40(2)$.
The results for all four DOS are summarized in the following table. 

\hspace{1cm}
\begin{center}
  \begin{tabular}{p{1.9cm}|p{1.5cm}|p{2cm}|p{2cm} p{0.001cm}}
    \centering DOS & \centering  $U_{c1}$ &  \centering  $r_{\hyb}$ &  \centering  $r_{\Sigma}$ & \\ \hline
    \centering semicircular & \centering $2.38$ & \centering $0.79 \pm 0.03$ & \centering $- 0.37 \pm 0.04$ & \\
    \centering narrow & \centering $1.43$ & \centering $0.77 \pm 0.03$ & \centering $- 0.4 \pm 0.02$  & \\
    \centering double & \centering $2.55$ & \centering $0.79 \pm 0.02$ & \centering $- 0.43 \pm 0.03$ & \\
    \centering hypercubic & \centering $2.36$ & \centering $0.87 \pm 0.04$ & \centering $- 0.36 \pm 0.03$ & \\
  \end{tabular}
\end{center}
\hspace{1cm}

The errors of the exponents were estimated from the confidence intervals of the numerical fits, the dependence of the exponents on the fitting interval and the uncertainty of $U_{c1}$ due to the difficult convergence close to the boundary of the coexistence region. The error bars (nearly) overlap, so no clear conclusion about the universality of the exponents is possible. The gap exponent $\nu z$ is in the range $0.77 \ldots 0.87$ for all four DOS. Thus, even if the exponents are non-universal, their dependence on the lattice is weak and their numerical values are robust.


\bibliographystyle{apsrev4-1}
\bibliography{references_publ}

\begin{thebibliography}{46}
\bibitem [{zer()}]{zerot_note}%
  \BibitemOpen
  \href@noop {} {}\bibinfo {note} {The zero-temperature Mott transition in
  DMFT, located at $U_{c2}$, has continuous (discontinuous) character when
  approached from the metallic (insulating) side, respectively. Scale-invariant
  behavior does not exist.}\BibitemShut {Stop}%
\bibitem [{gap()}]{gapcriterion_note}%
  \BibitemOpen
  \href@noop {} {}\bibinfo {note} {To study deviations from critical power-law
  behavior, a scale-invariant gap criterion like the one used here is
  required.}\BibitemShut {Stop}%
\bibitem [{ben()}]{benchmarkrho_note}%
  \BibitemOpen
  \href@noop {} {}\bibinfo {note} {We verified that our resistivity calculation
  reproduces the dc limit of the conductivity reported in
  Ref.~\onlinecite{BluemerDiss} for the metal, with deviations smaller than
  $25\%$ which are due to small differences in the self-energy arising from the
  use of different impurity solvers.}\BibitemShut {Stop}%
\end{thebibliography}

\end{document}